\documentclass[norwegian]{aa} 
\usepackage[hidelinks]{hyperref}
\usepackage{graphicx}
\usepackage{svg}
\usepackage{txfonts}
\usepackage{amsmath}
\usepackage{color}
\usepackage{bm}
\usepackage{natbib}
\usepackage{listings}
\usepackage{algorithmic}
\usepackage[normalem]{ulem}
\usepackage{sidecap}
\usepackage{placeins}
\sidecaptionvpos{figure}{t}

\makeatletter

\renewcommand*\aa@pageof{Page \thepage{} of \pageref*{LastPage}}

\makeatother

\def\be{ \begin{equation}}
\def\ee{ \end{equation} }

\newcommand{\diff}{{\mrm d}} 
\newcommand{\mrm}\mathrm 
 
\newcommand{\cyan}[1]{{\color{black}{#1}}}

\makeatletter
\newcommand*{\rom}[1]{\expandafter\@slowromancap\romannumeral #1@}
\makeatother
\begin{document} 


   \title{asimulation: Domain formation and impact on observables in resolved cosmological simulations of the (a)symmetron} 

   \author{\O{yvind} Christiansen
          \inst{1}\fnmsep\thanks{Corresponding author's \email{oyvind.christiansen@astro.uio.no}},
          Farbod Hassani\inst{1}, \and David F. Mota\inst{1}
          }

   \institute{Institute of Theoretical Astrophysics, University of Oslo,
             Sem S\ae{lands} vei 13, 0371 Oslo, Norway
             }

  \abstract
{ The symmetron is a dark energy and dark matter candidate that forms topological defects in the late-time universe and holds the promise of resolving some of the cosmological tensions.
We performed high-resolution simulations of the dynamical and non-linear (a)symmetron using the recently developed relativistic N-body code asevolution. By extensively testing the temporal and spatial convergence of domain decompositioning and domain wall stability, we determined criteria and physical intuition for the convergence. We applied the resolution criteria to {run} five high-resolution simulations with $1280^3$ {grids} and a box size of $500$ Mpc/h of the (a)symmetron{. W}e considered the behaviour of the scalar field and the domain walls in each scenario. We find the effect on the matter power spectra, the HMFs, and observables computed over the past light cone of an observer, such as the integrated Sachs-Wolfe and non-linear Rees-Sciama effect and the lensing, compared to $\Lambda$CDM. We show local oscillations of the fifth force strength and the formation of planar structures in the density field. The dynamics of the field was visualised in animations with high resolution in time. The simulation code is made publicly available.}

    \titlerunning{asimulation, resolved simulations of the (a)symmetron}
    \authorrunning{Christiansen et. al}
   \maketitle
%
\section{  Introduction}

The asevolution code \citep{christiansen_asevolution_2023} extends the relativistic N-body particle-mesh code gevolution \citep{adamek_gevolution_2016,daverio_latfield2_2016} by adding the (a)symmetron \citep{perivolaropoulos_gravitational_2022,hinterbichler_symmetron_2010}, which is a dynamical dark energy component. The theory is constructed so that the scalar field remains dormant during the early universe and undergoes a late-time phase transition that partitions the universe into domains of either the positive or negative field value minima. As the field takes on non-zero values, it starts mediating a fifth force that enhances the clustering of matter. The phenomenology of the (a)symmetron has been thought about in relation to cosmological tensions such as the Hubble tension \citep{perivolaropoulos_gravitational_2022,perivolaropoulos_challenges_2022,hogas_impact_2023} and cosmic dipole tension \citep{perivolaropoulos_isotropy_2023}, and it is plausibly connected to many more \citep{peebles_anomalies_2022,peebles_flat_2023,naik_dark_2022,nanograv_nanograv_2023}. The model has also been thought about as a candidate for dark matter \citep{burrage_radial_2017,burrage_symmetron_2019}; because of the environmental dependence of the mass of the field, the dark matter mass would correlate with the halo {and} galaxy masses, which has recently been shown to alleviate the core-radius scaling tension \citep{van_dissel_core_2023}. 
In the companion paper \citep{christiansen_gravitational_2024}, we simulate the production of stochastic gravitational waves by the (a)symmetron scalar field and indicate a region in the parameter space that may account for the observed spectral index of the stochastic gravitational wave signal as observed by the NANOGrav collaboration \citep{nanograv_nanograv_2023}. The symmetron screening mechanism, similarly to chameleon or Vainshtein screening, makes the fifth force inactive in local environments where the model would otherwise be subject to strict constraints from Solar System experiments \citep{bertotti_test_2003,esposito-farese_tests_2004,tsujikawa_constraints_2008}. Requiring the local Solar System or galaxy to be screened imposes some constraints on the allowed parameter space \citep{hinterbichler_symmetron_2010} that have recently been suggested, which were overestimated in the past due to the non-trivial interaction of the field with the environment \citep{burrage_accurate_2023}. We make asevolution publicly available to download\footnote{\cyan{\url{https://github.com/oyvach/asevolution}}}.

While we previously found convergence in results such as the background evolution and power spectra in \cite{christiansen_asevolution_2023}, we did not explore the convergence and resolution criteria for the domain wall network thoroughly. For the parameter choice and simulation settings we investigated, the agreement in the power spectrum between the dynamic and quasi-static solvers was indeed good, indicating that time-convergence of dynamics must be established using other quantities. Furthermore, the sizes of the simulations that formed stable domain walls were large, so that a systematic exploration of the convergence of the dynamic and non-linear part of the model would have been too expensive. Here, we set out to more rigorously establish the physical validity and convergence of the solutions found by our code. We do this in section \ref{S:convergence} by considering temporal and spatial resolution separately, and by constructing ideal situations that we take as representative of subsystems of the cosmological setup. We then move on to low-resolution cosmological simulations to find results on the convergence that agree with those of the ideal setups. Finally, in section \ref{S:hrsims}, we extrapolate our results to a suite of five high-resolution cosmological simulations. We visualise the behaviour of the scalar field in animations that we make available to the reader\footnote{\cyan{\url{https://oyvindchristiansen.com/projects/symmetron_resolve}}}.

\section{Conventions}\label{SS:units}
We set the speed of light $c$ equal to 1 and use the metric signature $(-++\,+)$. The conformal factor $A$ has an implicit dependence on the scalar field $\phi$. All symbols with a tilde over are defined in the Jordan frame, such as the trace of matter stress-energy tensor in the Jordan frame $\tilde T_m$; in the Einstein frame, it is written $T_m$.  By dot $\dot x$, we mean that the derivative is made with respect to cosmic time; primed quantities ($x'$) have their derivatives taken with respect to conformal time. We refer to the Compton wavelength corresponding to the mass $\mu$ as $L_C$, while the general Compton wavelength valid for all $T_m$ and spatial wavenumbers $k$ is referred to as $L$. We refer to $a_*$ as the scale factor at spontaneous symmetry breaking in a homogeneous universe, and $a_{\mrm{SSB}}$ as the scale factor of actual symmetry breaking. Finally, we refer to the `true vacuum' as the deepest minimum of the (a)symmetron potential when $T_m=0$, while we just write `vacuum' for the minimum otherwise, or `false vacuum'.

\section{Theory}
The asymmetron \citep{perivolaropoulos_gravitational_2022} is a generalisation of the symmetron \citep{hinterbichler_symmetron_2010}, both of which are defined with the action
 \begin{equation}\label{eq:action}
S  = \int \diff^{4} x \sqrt{-g} \, \Bigg[\frac{1}{2} M_{\text{pl}}^2 R-\frac{1}{2} \nabla_{\mu} \phi \nabla^{\mu} \phi-V(\phi)\Bigg]+ S_{m}\left(\psi_{i}, \tilde{g}_{\mu \nu}\right),
 \end{equation} 
where the integral is taken over spacetime, $g$ is the determinant of the Einstein frame metric $g_{\mu\nu}$, $R$ is the Ricci scalar corresponding to $g_{\mu\nu}$, $\nabla$ is the covariant derivative, $V(\phi)$ is the potential of the (a)symmetron scalar field $\phi$, and $S_m(\psi_i,\tilde g_{\mu\nu})$ is the matter action. $S_m$ is defined with the Jordan frame metric $\tilde g_{\mu\nu}$, but is otherwise as we know from the standard model of cosmology, consisting of cold dark matter (CDM), the cosmological constant, radiation, neutrinos, and baryons. The relation between the Jordan and Einstein frame metrics is given by the conformal transformation
 \begin{align}\label{eq:conformaltransform}
\tilde g_{\mu\nu}=A^2(\phi)g_{\mu\nu},
 \end{align} 
in which the conformal factor $A(\phi)$ is defined 
 \begin{align}\label{eq:conformalfactor}
A(\phi) \equiv 1 + \frac{1}{2}\left(\frac{\phi}{M}\right)^2 \equiv 1 +\Delta A,
 \end{align} 
where $M$ is the conformal coupling. Equation \ref{eq:conformaltransform} may be viewed as the attenuated Taylor expansion of some more generic function that is symmetric in $\phi$ as long as $\Delta A$ is small, which it is for the parameters that we consider in this paper. We can define the effective potential $V_{\mrm{eff}}$ that enters into the equations of motion of the (a)symmetron,
 \begin{equation}\label{eq:potential}
    V_{\text{eff}}  \equiv V - \ln A(\phi) T_m = V_0 - \frac{1}{2}\mu^2\phi^2 - \frac{1}{3}\kappa \phi^3 + \frac{1}{4}\lambda \phi^4 - \ln A(\phi) T_m,
 \end{equation} 
which has the additional term $\ln A(\phi) T_m$ that comes from the conformal transformation of the metric in the matter action (see e.g. \citep{christiansen_asevolution_2023}). $V_0$ is a constant, $\mu$ is the mass, $\lambda$ is the coupling constant, and $T_m$ is the trace of the Einstein frame stress-energy tensor. $\kappa$ is the coupling constant in the cubic term, and we recover the symmetron for the choice $\kappa=0$. We can define a set of phenomenological parameters that map to the Lagrangian parameters as in \cite{brax_systematic_2012,davis_structure_2012}, which was extended to the case of the asymmetron in \cite{christiansen_asevolution_2023}
 \begin{align}\label{eq:map1}
    L_{C} &= \frac{1}{\sqrt{2}\mu} \equiv \frac{\xi_*}{H_0} \approx \xi_* \cdot 2998 \text{ Mpc/h}, \\\label{eq:map2}
    M &= \xi_*\sqrt{6\Omega_{m,0} /a^{3}_*},\\\label{eq:map3}
    \lambda &= \left(\frac{M_{\mrm{pl}}}{M^2}\right)^2\frac{\mu^2}{\bar \beta^2 - \left(
    \Delta \beta/2
    \right)^2},\\
    \kappa &= \frac{\lambda M^2 \Delta \beta}{M_{\mrm{pl}}},
 \end{align} 
where $L_C$ is the Compton wavelength corresponding to the mass $\mu$ of the field, $H_0$ is the present time Hubble parameter, $\Omega_m$ is the present time background energy density parameter for matter, $M_{\mrm{pl}}=1/\sqrt{8\pi G_N}$ is the Planck mass, and $G_N$ is Newton's gravitational constant. The Lagrangian parameters are then defined in terms of $(\xi_*,z_*,\beta_+,\beta_-)$, where $\xi_*=L_C H_0$ is the ratio of the Compton wavelength to the present time Hubble horizon, $z_*$ is the cosmological redshift at symmetry breaking for a homogeneous universe, and $\beta_{\pm}$ are the force strengths relative to the Newtonian force in the positive and negative symmetry-broken minima. We also defined $\Delta\beta = \beta_+-\beta_-$ and $\bar\beta = (\beta_++\beta_-)/2$. The field equations can be found using the Euler-Lagrange equation, and they can be set in the form \citep{llinares_releasing_2013} when neglecting relativistic corrections, 
 \begin{align} \label{eq:ELeq1}
    \dot q &= a \nabla^2\chi -a^3 \mu^2\left\{
    \chi^3 - \bar \kappa \chi^2 + \left(
    \eta  - 1
    \right) \chi
    \right\},\\
    q &= a^3 \dot \chi,\quad \chi = \phi/v,
 \end{align} 
where $v = \mu/\sqrt{\lambda}$ is the true vacuum expectation value of the scalar field in the symmetron scenario. The extension of the equations of motion to include relativistic corrections is provided in \cite{christiansen_asevolution_2023}, but is not considered for the purposes of the present consideration. 

\section{Convergence} \label{S:convergence}

In this section, we review our results on the convergence of the field solver in finding the domain wall network. 
In \cite{christiansen_asevolution_2023}, we made use of the leapfrog solver to evolve the scalar field. During convergence testing, we initially found a high degree of sensitivity of the domain wall network to initial conditions as well as to changes in both temporal and spatial resolution. Therefore, we found it appropriate to first investigate a small selection of solvers to determine whether a different field solver might perform better. This made us choose the explicit fourth-order Runge-Kutta solver in the following. We refer to appendix \ref{A:solvers} for a review of the different solvers and their convergence. A useful property of the explicit Runge-Kutta solver that we make use of here is that it becomes unstable and diverges after accumulating a sufficient amount of error. Implicit schemes have larger stability regions and can provide (sometimes unconditionally) stable and reasonable solutions even when the physical system is not being resolved \citep{linge_diffusion_2017}. We use the stability of the explicit scheme as an indicator for whether the physical solution is being resolved, which is motivated by our ideal system considerations in section \ref{SS:timescale}. We first review the relevant scales of the model. Next, we describe the convergence of physical solutions during the time after symmetry breaking. We use both idealised situations and realistic cosmological cases, where we also consider the initial conditions with care. Unless otherwise specified, we use in all instances below the symmetron parameters $(L_C,z_*,\beta_*) = (1\,\mrm{Mpc/h}, 2, 1)$ that were also studied in \cite{llinares_cosmological_2014,christiansen_asevolution_2023}. In this section, we refer to the Courant factor as the factor defined
 \begin{align}\label{eq:Courant}
    C_f = v_{\mathrm{vel}}\, \mrm{d} \tau / \mrm{d} x \equiv v C,
 \end{align} 
where for the scalar field $C_{f,\phi}=C_\phi$, we set $v^{(\phi)}_{\mrm{vel}}=1$, and $\mrm d \tau \rightarrow \mrm d \tau_\phi$; $\tau$ is the conformal time coordinate. On the other hand, for CDM, $C_{f,\mrm{cdm}}$, we use $v_{\mrm{vel}}=v_{\mrm{cdm,max}}\sim 0.02$ and $\mrm d \tau \rightarrow \mrm d \tau_{\mrm{cdm}}$. In other words, we vary the time stepping of CDM and the scalar field separately, requiring $\mrm d \tau_{\phi}\leq \mrm d \tau_{\mrm{cdm}}$. For convenience, we show the time convergence for the CDM equations in terms of $C_{\mrm{cdm}}=\mrm{d}\tau/\mrm{d}x$.

\subsection{Relevant scales}\label{SS:timescale}
\begin{figure}[ht]
    \centering
    \includegraphics[width=\linewidth]{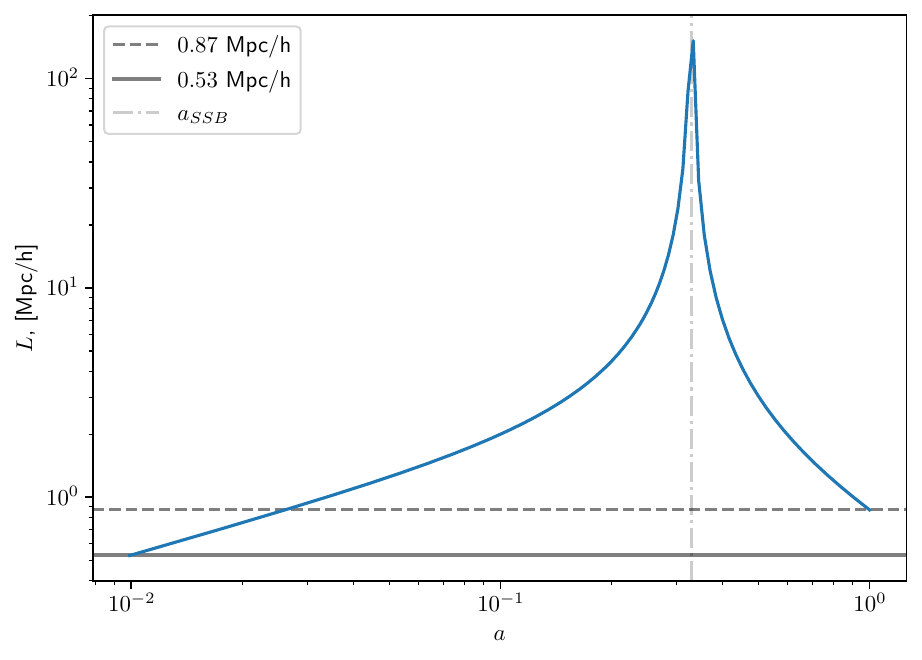}
    \caption{General Compton wavelength of the symmetron field at the background as a function of the scale factor. The parameter choice is $(L_C,a_*,\beta_*) = (1\text{ Mpc/h}, 0.33,1)$, and the horizontal lines indicate the initial and final Compton scales }
    \label{fig:timeresolution_rule}
\end{figure}

In order to reliably evolve the scalar field, the frequency scale of its oscillations should be smaller than the Nyquist wavenumber of our solver. In
\cite{christiansen_asevolution_2023}, we found the linear solution of the perturbed symmetron in Fourier space, which is a damped oscillation with phase frequency $\omega$, so that
 \begin{align} \label{eq:analyticfrequency}
    \omega^2 = k^2 + a^2m^2 - \mathcal{H}^2 \equiv \frac{1}{2 L^2},
 \end{align} 
where the mass $m$ is defined below, $\mathcal{H} = a H$ is the comoving Hubble parameter, $k$ is the comoving wave number, and we defined the general Compton scale $L$.   
 For large scales $k L\ll 1$, for a subhorizon general Compton wavelength $L H \ll 1$, the relevant conformal timescale is
 \begin{align} \label{eq:generalCompton}
    L = \frac{1}{\sqrt{2} a m }.
 \end{align} 
The mass $m$ above makes use of the definition $m^2 = \left | \partial^2_\phi V \right |_{\phi=v}$, which gives 
 \begin{align} \label{eq:generalMass}
\text{before SSB, \;\;} \left|T_m\right| \geq \rho_*: m&= \mu \sqrt{-\frac{T_m}{\rho_*}-1 },\\
 \text{after SSB, \;\;}   \left|T_m\right| < \rho_* : m&= \sqrt{2}\mu\sqrt{ 1+T_m/\rho_*}, 
 \end{align} 
where $\rho_*\equiv \mu^2 M^2$ is the energy scale of the symmetry breaking. On the background, when $\Omega_\phi \ll \Omega_m$ and $\Delta A\ll 1$, $T_m/\rho_*\simeq -\left(a_*/a\right)^3$.
 For the fiducial symmetron model choice $(L_C,z_*,\beta_*) = (1\text{ Mpc/h}, 2,1)$, we see the background evolution of the conformal Compton scale shown in figure \ref{fig:timeresolution_rule}.
We demonstrate more clearly in sections \ref{SS:spaceresolution} and \ref{SS:cosmologyconvergence} that this scale relates to whether we are resolving the system. Taking this for granted for now, we show in figure \ref{fig:timeresolution_rule} that the timescale of the oscillations is initially set very small due to the high mass gained from the background energy density. It is therefore very expensive to evolve the field reliably from $z\sim 100$ until symmetry breaking, for instance. The mass $m$ goes to zero around the time of the symmetry breaking. From expression \eqref{eq:analyticfrequency}, the Hubble horizon is then $ k^2-\mathcal{H}^2 \sim \frac{1}{2 L^2}\rightarrow 0$, which is often thought to set the scale of percolation along with causality, as discussed for example in \cite{llinares_domain_2014}.
\subsection{Courant condition}\label{SS:courantfactor}
\begin{SCfigure*}
    \centering
    \includegraphics[width=1.5\linewidth]{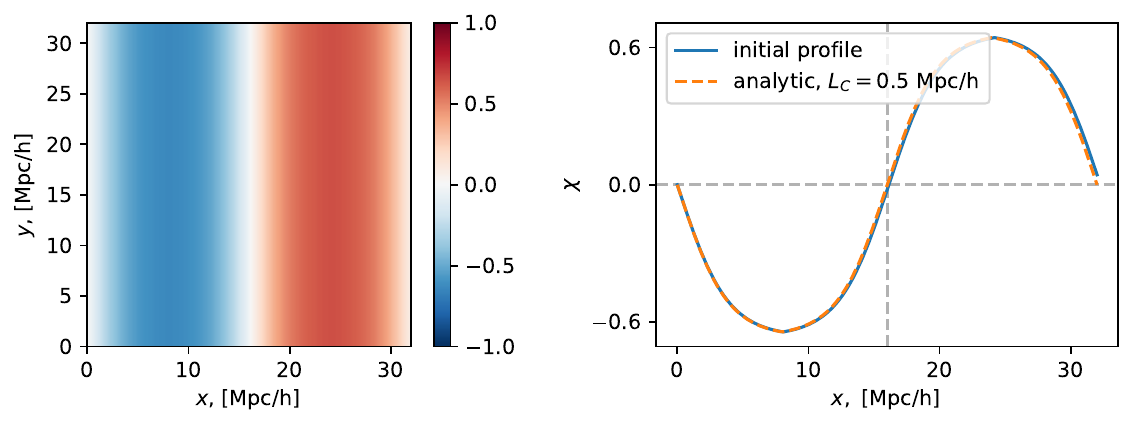}
    \caption{
    \protect\rule{0ex}{5ex}
    Initial setup for the bare wall spatial and temporal convergence experiments. The right panel shows the overlaid analytic profile corresponding to a Compton wavelength of $0.5$ Mpc/h, although the model here is defined with a Compton wavelength of $1$ Mpc/h.}
\label{fig:idealgriddxconvergence_setup}
\end{SCfigure*}

\begin{SCfigure*}
    \centering
    \includegraphics[width=1.3\linewidth]{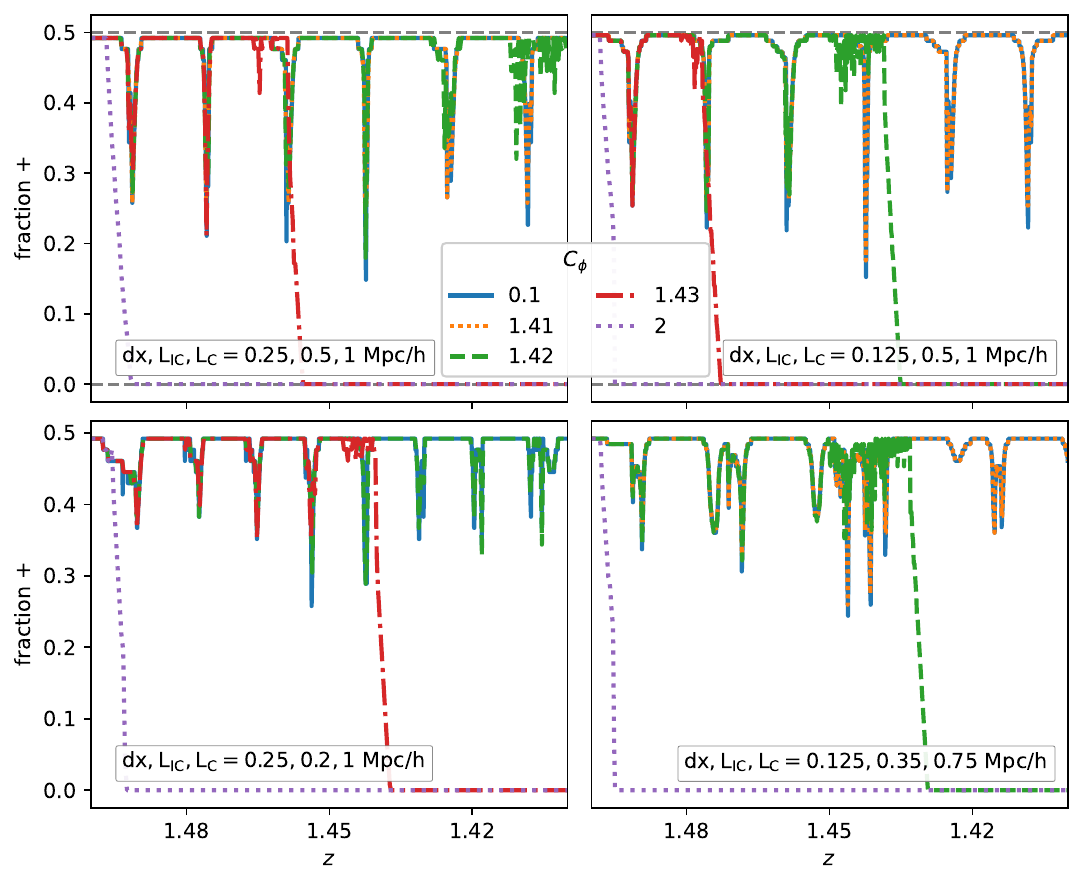}
    \caption{
    \protect\rule{0ex}{5ex}
    Volume fraction of the field that is in the positive minimum as a function of redshift for the oscillating bare wall system for variations of the Courant factor $C_\phi$. The four plots vary in spatial resolution $\mrm d x$, initial wall scale $L_{\mrm{IC}}$, and Compton wavelength $L_C$. }
    \label{fig:idealCaseOscillationTime}
\end{SCfigure*}

The section \ref{SS:timescale} indicates the resolution scale we expect to need in order to resolve the symmetron on the background. When the field is resolved, we still expect a condition on the Courant factor, which comes from the stability of the numerical scheme. We determine this condition next. We constructed an idealised system where the matter field is homogeneous and evolves as $\rho_m = \rho_{m,0}/a^3$. We initialised the scalar field with its expectation value for a homogeneous background $v$ and with a sign difference $\pm v$ on the different halves of the box, so that there was a centre domain wall. At first, we ran the Gaussian relaxation iterations presented in \cite{llinares_isis_2014,christiansen_asevolution_2023} to thermalise the field configuration and smooth the steep gradient at the centre. The higher the tolerance we used for the relaxation, the closer the configuration of the field became to the analytic profile. If it had exactly the analytic profile, the field was unexcited and stayed dormant. If the box was too small for the analytic field profile to establish, the relaxation instead placed the field entirely at one minimum, decided by the numerical error. Since we are interested in considering a dynamic scenario, we set the relaxation tolerance to not relax the field entirely. In the right panel of figure \ref{fig:idealgriddxconvergence_setup}, we show the initial profile of the field. Overlaid is the analytic profile corresponding to a smaller Compton wavelength of $0.5 $ Mpc/h, although the field we used here had a Compton wavelength of $L_C = 1$ Mpc/h. The field {was} then in an excited state, where the profile was initially squeezed and started to oscillate when it {was} evolved.
In figure \ref{fig:idealCaseOscillationTime} we show the evolution of a fraction of lattice points in the positive minimum for different choices of temporal and spatial resolutions, in addition to different Compton and initial scales. We take the convergence of this fraction to be a proxy for whether the field configuration is properly resolved. To gain more control of the initial conditions as we compare different spatial resolutions, we simply initialised the field in the analytical profile corresponding to a smaller Compton wavelength of $L_{\mrm{IC}}=0.5$ Mpc/h. Because we used the explicit Runge-Kutta solver, the solutions diverged immediately when they accumulated a sufficient amount of error, and this indicated the critical Courant factor for stability of the numerical scheme, which we identify here to be $C_{\phi,\mrm{crit}}\sim 1.4$. This means that we require $\mrm{d}\tau \le 1.4\, \mrm{d}x$. As long as the Compton wavelength is resolved, the critical Courant seems to be independent of $\mrm d x, \,L_{\mrm{IC}},$ and $L_C$.

\subsection{ Spatial resolution}\label{SS:spaceresolution}
\begin{figure}[ht]
    \centering
    \includegraphics[width=\linewidth]{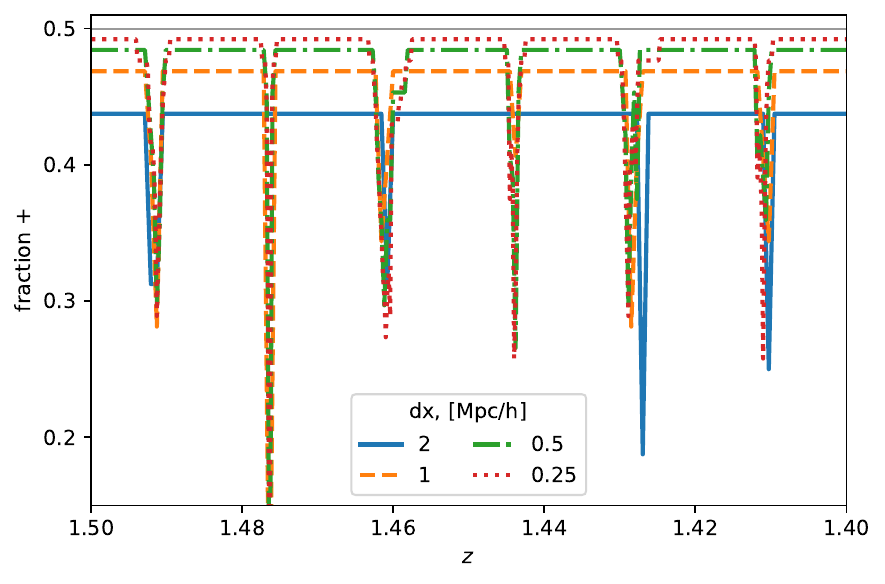}
    \caption{Volume fraction of the field in the positive minimum as a function of redshift for the oscillating bare wall system for a selection of spatial resolutions. The coarsest simulation has the Courant factor choice $C_\phi=0.5$ and would otherwise not resolve the oscillations. The rest are kept with $C_\phi=1$.}
    \label{fig:idealCaseSpatialRes}
\end{figure}
The smallest spatial scale introduced by the symmetron is the interface between two different domains. The profile of the field along an axis, $x$, orthogonal to the domain wall looks like what is shown in the right panel of figure \ref{fig:idealgriddxconvergence_setup}. As was shown in \cite{llinares_domain_2014}, for example, the analytic expression for a domain wall in a homogeneous matter distribution is given by
 \begin{align} \label{eq:barewall}
    \chi (x,a) = \sqrt{1-\left(\frac{a_*}{a}\right)^3}
    \tanh \left(
    \frac{a x}{2 L_C}\sqrt{1-\left(\frac{a_*}{a}\right)^3}
    \right),
 \end{align} 
where $L_C$ is the Compton wavelength corresponding to the Lagrangian mass parameter $\mu$, as shown in equation \eqref{eq:map1}.
In figure \ref{fig:idealCaseSpatialRes}, we show the oscillation in a fraction of lattice points in the positive minimum for different spatial resolutions, keeping the initial wall and Courant scales $L_{\mrm{IC}},L_C=0.5,1$ constant. In each case, we fixed the Courant factor $C_\phi = 1 < 1.4$ according to the numerical stability found in the previous subsection, \ref{SS:courantfactor}. For a spatial resolution of $\mrm d x=2$ Mpc/h, which is higher than the general Compton resolution, we saw an aliasing of the correct oscillation. In figure \ref{fig:idealCaseSpatialRes} we instead used a Courant factor $C_\phi=0.5$ for the case when $\mrm d x = 2$ Mpc/h to determine whether the spatial resolution is important by itself. In this case, some of the fluctuations were captured, but the second and third beats were missed while the fourth was slightly displaced. Our conclusion then is that the spatial resolution itself has to be smaller than $L$, which in turn is $L\geq L_C/\sqrt{2}$ in the symmetry-broken phase. For still lower spatial resolutions, the gradients are increasingly well resolved, and the oscillation is smoother. 

\subsection{Convergence in  cosmological setups}\label{SS:cosmologyconvergence}
\begin{SCfigure*}
    \centering
    \includegraphics[width=1.5\linewidth]{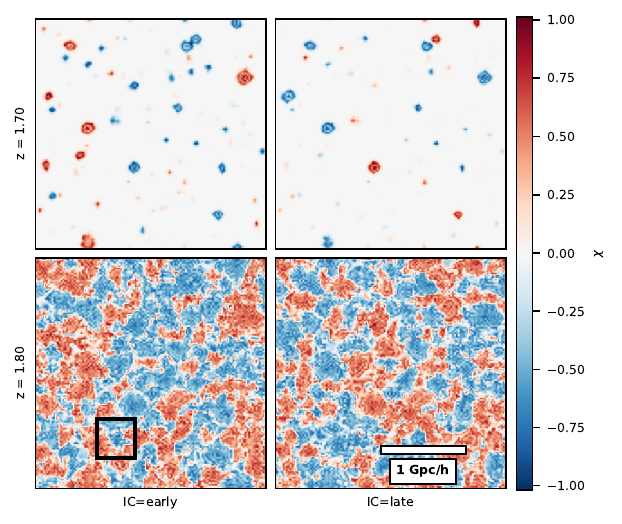}
    \caption{
    \protect\rule{0ex}{5ex}
    Comparison of snapshots of the scalar field $\chi$ at redshifts $z=1.9$ and $1.8$ for a late initialisation at $z=2.5$ and for an early initialisation at $z=100$. The amplitude of the late-time initial conditions is chosen to be similar to the evolved early-time conditions. In the bottom left snapshot, a black square indicates the relative size of a $\left(500\,\mrm{Mpc/h}\right)^3$ volume box.}
    \label{fig:ic_comparison}
\end{SCfigure*}
In the above section \ref{SS:courantfactor}, we found the condition for numerical stability, and in \ref{SS:timescale} and \ref{SS:spaceresolution}, we indicated the relevant scale to resolve the physical system. We applied these criteria to a cosmological scenario. Some complications to the previous ideal situation arose: First, the full set of partial differential equations that we considered was extended by the dark matter and metric perturbations, which are no longer treated at the background. In principle, this may affect the numerical stability. Next, since the matter fields are not treated homogeneously, both regions of the field are in the symmetry-broken and -unbroken phases. In the unbroken symmetry, for large $\rho_m$, $L\sim M/\sqrt{\rho_m}$ may be smaller than $L_C$, and it may be important to resolve if {we want} convergence on the exact field configuration of the scalar is to be achieved. In the following, we ensured to resolve the Compton scale of the symmetry-broken phase by fixing $\mrm d x=0.8$ Mpc/h and $C_\phi \lesssim 1$.

We initialised the scalar field from a scale-invariant power spectrum at $z=100$ for a demonstration: When the field is evolved at a time resolution that is too coarse, for example $0.7$ Mpc/h, when the general Compton at the background at initialisation time is $0.53$ Mpc/h, the realisations of the domain wall network are completely different in general when the time-stepping is varied slightly. The evolution in the pre-symmetry-broken phase is not very interesting for our parameter choices, however, and the power spectrum of the field remains close to scale-invariant right up until the symmetry breaking (see appendix \ref{A:initialisation}{)}. We demonstrate in figure \ref{fig:ic_comparison} that we obtain a qualitatively similar domain decompositioning when we instead initialise the scale-invariant power spectrum at some short duration before the symmetry breaking, where we used a very coarse spatial resolution and the more stable leapfrog solver for demonstration. A late initialisation amounts to a large optimisation of the computational cost.

By evolving the scalar field in a cosmological scenario together with the CDM and metric perturbations, using the resolution criteria developed in the ideal-case bare wall scenario, we find solutions that blow up and diverge. This indicates that not all relevant scales were resolved; in the current setup, we used a spatial resolution of $\mrm d x = 0.8$ Mpc/h and a Courant factor of the scalar $C_\phi \sim 1<1.4$. We expected the stability to change both because new relevant scales were introduced by the density contrasts that maintained the $\mathbb{Z}_2$ symmetry in high-density areas and because the numerical scheme changed from co-evolving CDM, metric perturbations, and the scalar field. Because we varied $\mrm d \tau_\phi$ separately, we expected the numerical scheme to be similar to the ideal case when the other dynamical degrees of freedom evolved slowly in comparison, which we expected them to. The new minimum scale in the symmetric phase owing to the density contrasts is
 \begin{align}
    L^{\mrm{min}} \sim \frac{L_C}{\sqrt{ a_*^3}}\sqrt{\frac{a}{\delta_m^{\mrm{max}}}}.
 \end{align} 
For the spatial resolution that we considered in the high-resolution simulations, $\mrm d x\sim 0.4$ Mpc/h, $\delta_m^{\mrm{max}}\sim 3000$, meaning that conservatively, $L^{\mrm{min}}\sim 0.013$ Mpc/h. In other words, we did not expect to resolve the false vacuum fluctuations of the field in the highest-density environments for computationally reasonable resolutions, and we  therefore expect some small degree of phase randomisation. We were still able to avoid divergence of the solutions by adjusting the Courant factor to $C_\phi \sim 0.15$. In this event, we resolved the Compton scale in the symmetric phase for density contrasts $\delta_m\lesssim 140$ in the event of $\mrm d x\sim 0.4$ Mpc/h, which was satisfied by almost the entire simulation box volume. For the choice $\mrm d x\sim 0.8$ Mpc/h, it is instead sufficient to use $C_\phi\sim 0.2$. As in the ideal case, in subsection \ref{SS:courantfactor}, we expect the non-divergence of the solutions to indicate that the system is sufficiently resolved.
\begin{SCfigure*}
    \centering
    \includegraphics[width=1.5\linewidth]{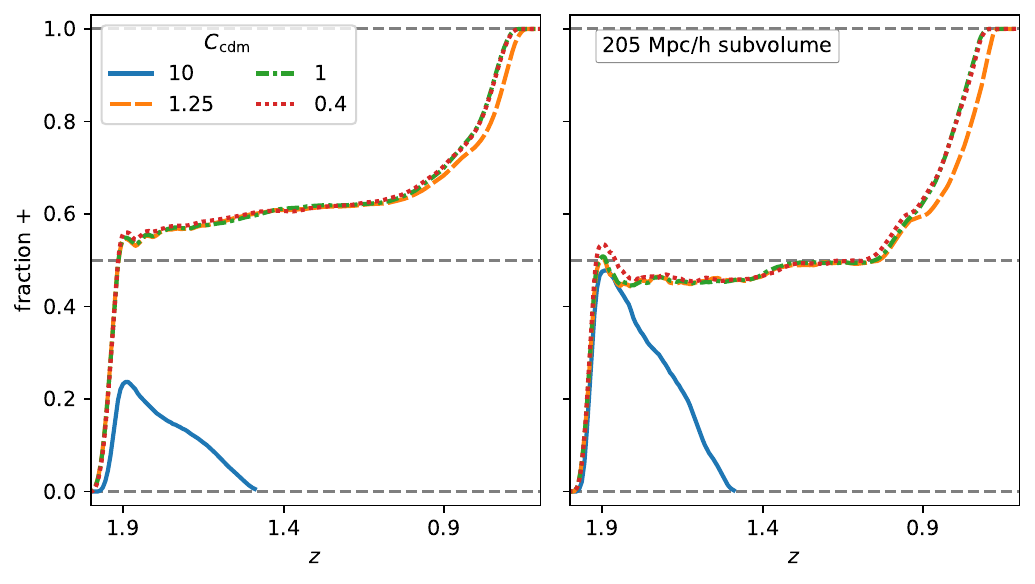}
    \caption{
    \protect\rule{0ex}{5ex}
    Fraction of nodes in positive minimum for simulations run with initial conditions found from a $\Lambda$CDM simulation with a Courant factor $C_{\mrm{cdm}}=1$. The box size is $B=410$ Mpc/h, and to the right, we show the fraction of domains within a $B=205$ Mpc/h subvolume. We used $512^3$ grids. We chose $C_\phi=0.15$.}
\label{fig:CosmologicalCaseDomainFraction}
\end{SCfigure*}

In the case of $\Lambda$CDM, the timescale of the evolution for the CDM is much longer than what we considered here for the symmetron. For large simulations, Courant factors can typically be used in gevolution so that $C_{\mrm {cdm}} = \mrm d t/\mrm d x \sim 50$. Since the CDM particles are slow, for the type of resolution we considered, $\mrm d x\sim 0.4 $ Mpc/h, the maximum velocity is $v_{\mrm{cdm}}^{\mrm{max}}\sim 0.02$, which means $C_{f,\mrm{cdm}}\sim 1$ in this instance. However, due to the interaction of the symmetron and the dark matter, {additional} dynamics of the CDM is introduced on the symmetron timescale. As an intuitive requirement for resolving this dynamic, we suggest $v_{\mrm{cdm}}^{\mrm{max}}\mrm d\tau_{\mrm{cdm}}\ll L(T=0)\leq L_C / \sqrt{2}$. The fifth force is screened when the symmetry is restored, so that the CDM time-stepping only needs to resolve the symmetry-broken Compton scale. For $\mrm d x\sim0.4$ Mpc/h, we then used a Courant factor of $C_{\mrm{cdm}}=1$, while for $\mrm d x \sim0.8$ Mpc/h, we chose $C_{\mrm{cdm}}=0.4$. In both instances, the maximum displacement of a dark matter particle during one iteration is $\Delta x_{\mrm{cdm}}^{\mrm{max}}\sim 0.4 \cdot  v_{\mrm{cdm}}^{\mrm{max}} \,\mrm{Mpc/h}\sim 0.01$ Mpc/h $\sim $ 1 \% of the scale of the domain walls. In figure \ref{fig:CosmologicalCaseDomainFraction} we show convergence in the fraction of lattice points in the positive symmetry-broken minimum as a function of redshift for decreasing Courant factors and for a spatial resolution of $\mrm d x \sim 0.8$ Mpc/h. There is approximate convergence for $C_{\mrm{cdm}}\sim 1$.

\section{High-resolution simulations} \label{S:hrsims}
In the previous section, \ref{S:convergence}, we established resolution criteria, guided by physical intuition, ideal system simulations, and cosmological system convergence testing. In this section, we apply them to a suite of five $500$ Mpc/h, $1280^3$ grid simulations in order to predict the non-linear and and dynamically dependent observables that are not accessible by use of the non-converged, semi-analytic, or quasi-static approaches used in the past.
\begin{table}[ht]
    \centering
    \begin{tabular}{|c|c|c|c|c|}\hline
        model & $L_C$ [Mpc/h] & $z_*$ & $\beta_+$ & $\beta_-$  \\\hline
        \rom{1} & 1 & 2 & 1 & 1 \\
        \rom{2} & 1 & 0.1 & 1 & 1\\
        \rom{3} & 1 & 0.1 & 8 & 8 \\
        \rom{4} & 0.75 & 0.1 & 8 & 8 \\
        \rom{5} & 1 & 0.1 & 8.4 & 7.6 \\\hline
        \multicolumn{5}{c}{}
    \end{tabular}
    \caption{Model with five different parameter choices \rom{1}-\rom{5}. $L_C$ is the Compton wavelength corresponding to the mass $\mu$, $z_*$ is the redshift of the symmetry breaking in a homogeneous universe, and $\beta_+,\beta_-$ are the fifth-force strengths in the positive and negative field-value minima, respectively. }
    \label{tab:modelparameters}
\end{table}
\begin{figure}[ht]
    \centering
    \includegraphics[width=\linewidth]{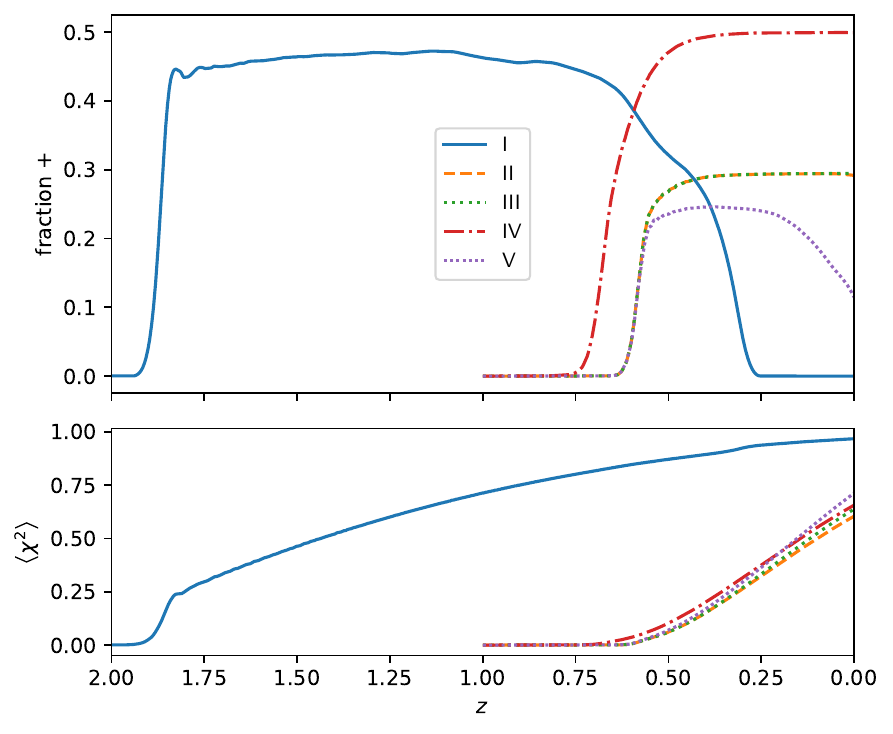}
    \caption{{Overview of background quantities for the suite of simulations.} Top: Fraction of the simulation volume where the scalar field occupies the positive minimum for the five different models \rom{1}-\rom{5}. Bottom: Average over the simulation volume of the square of the scalar field $\chi^2$. Both quantities are plotted as a function of the redshift $z$.}
    \label{fig:background_overview}
\end{figure}

In table \ref{tab:modelparameters} the (a)symmetron parameters are shown for each of the simulations \rom{1}-\rom{5}. The cosmological parameters are otherwise as found by the Planck satellite \citep{planck_planck_2020}, with the Hubble parameter $h=0.674$, and the cosmological energy density parameters at $z=0$ are $\Omega_b,\Omega_m,\Omega_\Lambda=0.049,0.264,0.687$. We adjusted $\Omega_\Lambda$ to satisfy the flatness criterion $\sum_i \Omega_i=1$, although the symmetron parameter choices considered have $\Omega_\phi\lesssim 10^{-5}$, so that the adjustment is very small. We used the cosmic microwave background temperature $T_{\mrm{cmb}}=2.7255$ K, the spectral amplitude and index $A_s,n_s=2.1\cdot 10^{-9},0.965$, and the spectral tilt $k_p=0.05 \, \mrm{Mpc}^{-1}$. For simplicity, we did not consider massive neutrinos and set the effective number of relativistic degrees of freedom $N_{\mrm{eff}}=3.046$. We chose the number of CDM particles equal to the number of grids.
\subsection{Description of the data}
\begin{figure*}[ht]
    \centering
    \includegraphics[width=\linewidth]{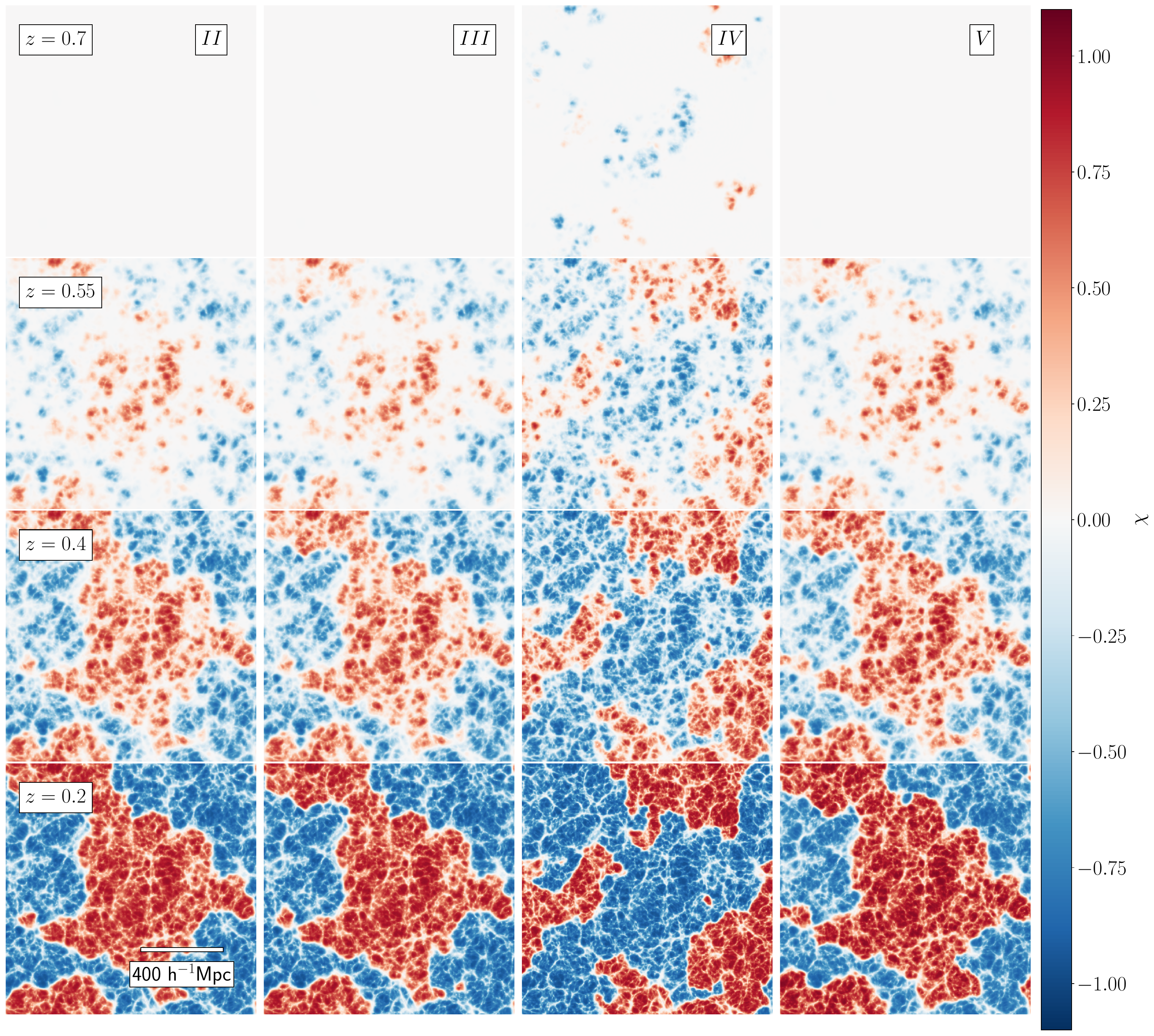}
    \caption{Snapshots of the scalar field $\chi${. Shown for} models \rom{2}-\rom{5} from left to right, for four different redshifts from top to bottom. }
    \label{fig:field_cartoon}
\end{figure*}

\begin{figure*}[ht]
    \centering
    \includegraphics[width=\linewidth]{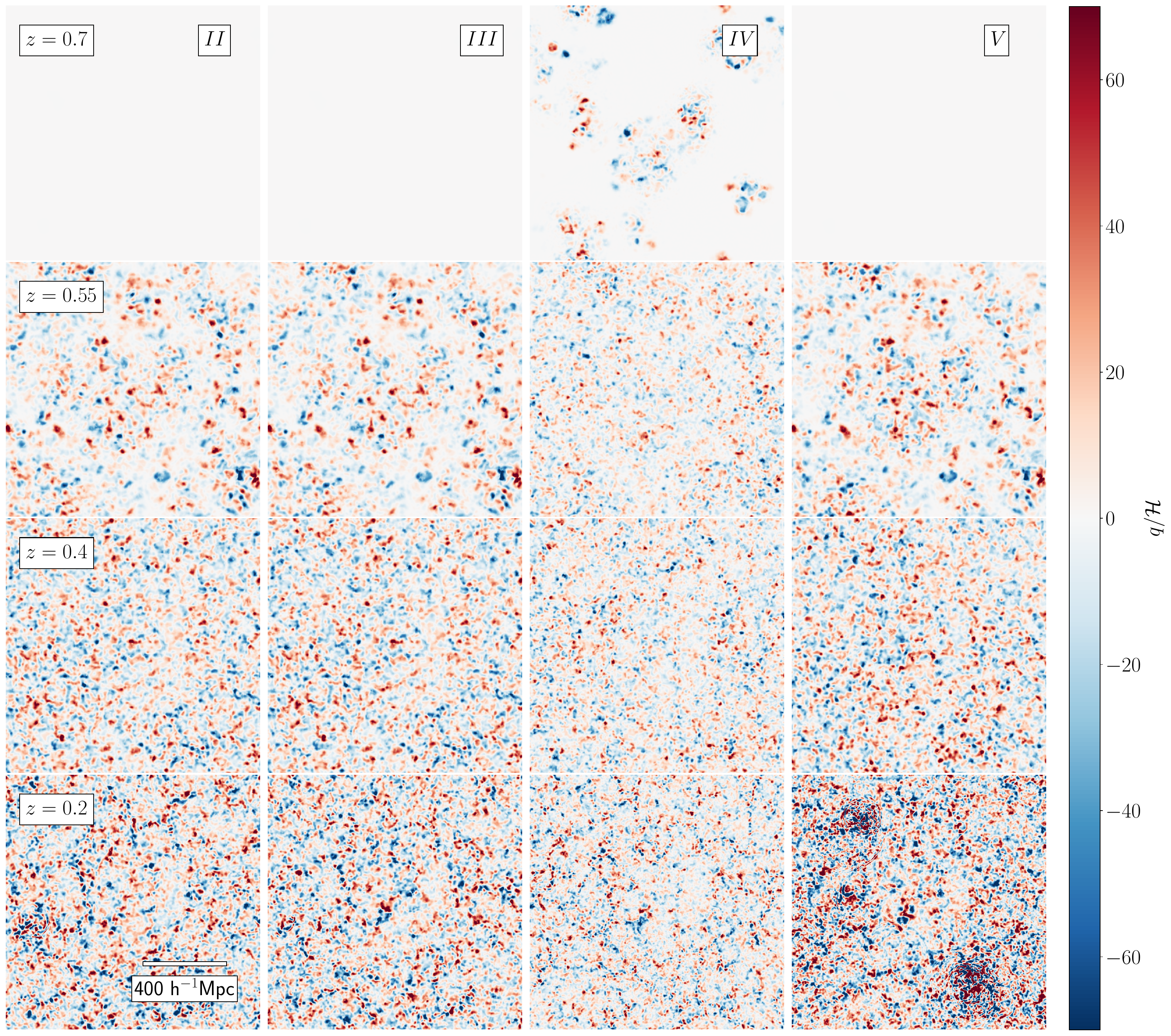}
    \caption{Snapshots of the velocity of the scalar field $q=a^3 \dot \chi${. Shown for} models \rom{2}-\rom{5} from left to right, for four different redshifts from top to bottom. The field is normalised to the conformal Hubble parameter. }
    \label{fig:aq_cartoon}
\end{figure*}

\begin{figure*}[ht]
    \centering
    \includegraphics[width=\linewidth]{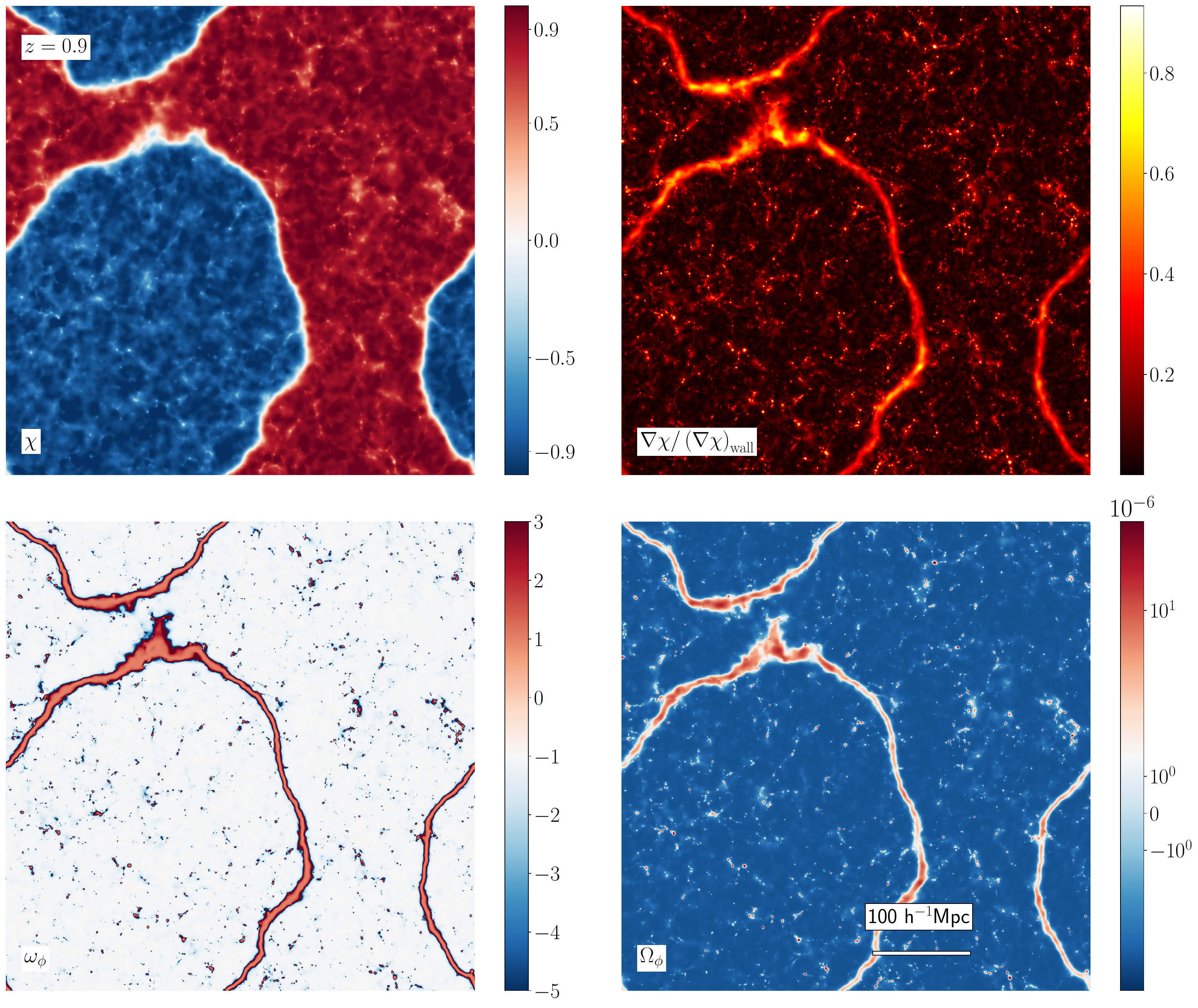}
    \caption{Snapshots {of slices of the simulation volume.} {F}rom top left to bottom right: The scalar field $\chi$, the isotropic $\nabla \chi\equiv$ $\left | \bf{\nabla}\chi \right |/\sqrt{3}$ normalised to the one expected for the centre of the analytic bare wall, the equation-of-state parameter of the field $\omega_\phi$, and the energy density parameter $\Omega_\phi=\rho_\phi/\rho_c$. The snapshots are from model \rom{1} and are shown at redshift $z=0.9$ at a critical moment right before the two blue domains join.}
    \label{fig:4plot}
\end{figure*}

\begin{figure*}[ht]
    \centering
    \includegraphics[width=\linewidth]{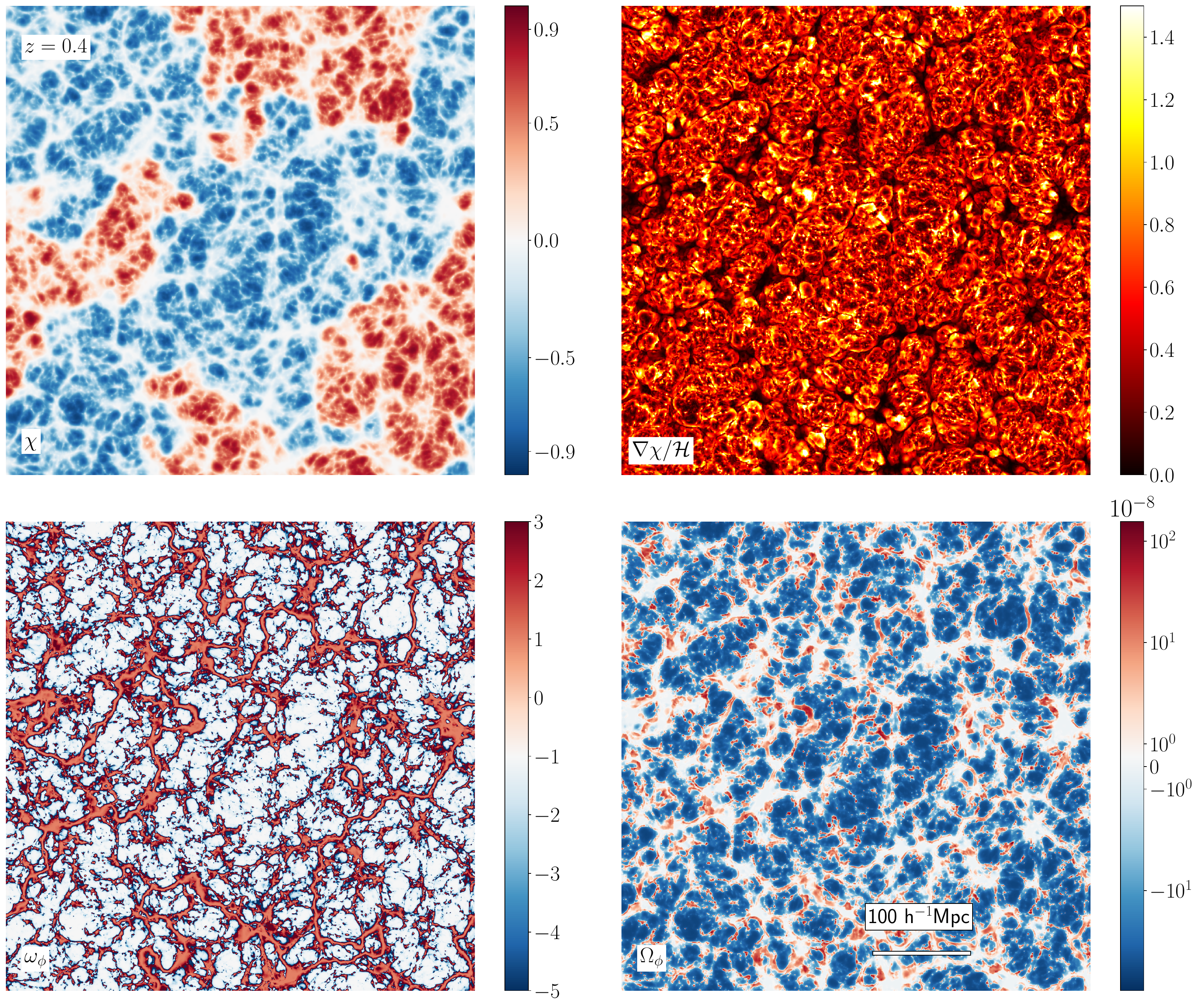}
    \caption{Snapshots {of the simulation volume.} {F}rom top left to bottom right: The scalar field $\chi$, the isotropic gradient $\nabla\chi\equiv${$\left | \bf{\nabla}\chi \right |/\sqrt{3}$} normalised to the conformal Hubble parameter $\mathcal{H}$, the equation-of-state parameter of the field $\omega_\phi$, and the energy density parameter $\Omega_\phi=\rho_\phi/\rho_c$. The snapshots are from model \rom{4}.}
    \label{fig:4plot2}
\end{figure*}

In figure \ref{fig:background_overview} we give an overview of the symmetry-breaking process for the different parameter choices, illustrated by the fraction of the simulation volume in the positive field-value minimum and the average of the square of the field as a function of redshift. We comment on some interesting features  {of} the background:
\begin{enumerate}
    \item The domain wall for the early symmetry-breaking scenario, model \rom{1}, is only quasi-stable and collapses at redshift $z\sim 0.5$. 
    \item The asymmetron scenario, model \rom{5}, is only quasi-stable and undergoes a slow collapse from redshift $z\sim 0.25$.
    \item All three instances of symmetron $\Delta\beta=0$ late-time symmetry breaking $z_*=0.1$, models \rom{2}-\rom{4}, seem to be stable.
    \item While the timescale from the symmetry breaking of the first voids to the last voids, that is, the width of the transition, is rather small for these parameter choices, it grows slightly longer for smaller $L_C$.
    \item The timescale for the minima to reach the true vacuum, indicated by $\langle \chi^2\rangle $, is comparatively long, so that the vacua still evolve throughout the simulation time.
    \item The bump in $\langle \chi^2\rangle$ at the onset of the symmetry breaking for model \rom{1} bears witness of an initially non-adiabatic evolution where the field falls into the minimum. The phase transition seems to occur more smoothly in the late-time symmetry breaking, models \rom{2}-\rom{5}.
\end{enumerate}
We discuss the observational implications in section \ref{S:discussionsandconclusion}.
In figures \ref{fig:field_cartoon} and \ref{fig:aq_cartoon} we give an overview of the evolution of the exact field configuration as viewed through snapshot slices through the simulation volume at four different redshifts.
\begin{enumerate}
    \item In addition to the earlier symmetry breaking in the smaller Compton wavelength model, \rom{4}, a different domain partitioning occurs in this instance as a result of the different field phases at the time of symmetry breaking. We expect this both from the different oscillation timescale of the field $\sim L$ and from the different symmetry-breaking time.
    \item Variation in the coupling $\beta$ shows no strong effects, but some small collapses in the weaker coupling case, model \rom{2}, indicate that the domain walls are less strongly pinned to filaments in this instance.
    \item Some of the domain wall collapses are especially clear in figure \ref{fig:aq_cartoon}, where the velocity shows large amplitude oscillations around the site of the collapse.  {For example, the dark regions in the bottom right and top left corner of the $z=0.2$ plot in the fourth column (labelled \rom{5}) of figure \ref{fig:aq_cartoon} correspond to regions where the domain wall has moved between $z=0.4$ and $z=0.2$ in figure \ref{fig:field_cartoon}. }
    \item Otherwise, we note that the velocity field correlates on the scale of the Compton wavelength $\sim L$, being on smaller scales in the instance of model \rom{4}.
\end{enumerate}
Finally, in figures \ref{fig:4plot} and \ref{fig:4plot2}, we show the scalar configuration at an instance for models \rom{1} and \rom{4}, respectively. We show the same simulation slice in four different fields: The scalar field $\chi$; the isotropic gradient $\left| \vec \nabla \chi \right|/\sqrt{3}$, normalised to the bare wall analytic expectation at the centre, $\left| \vec \nabla \chi \right|_{\rm wall} = \frac{\sqrt{3} a }{2 L_C} \left( 1-(\frac{a_*}{a})^3 \right) $, in the case of model \rom{1} and to the Hubble horizon $\mathcal{H}$ in the case of model \rom{2}-\rom{4}, where the analytic expectation is not available since $a<a_*$; 
the equation of state of the field, $\omega_\phi$ in the rest frame of the simulation; and finally, the energy density parameter of the field $\Omega_\phi = \rho_\phi/\rho_c$, where $\rho_c=3 H^2 M_{\mrm{pl}}^2$ is the critical density. Our comments are listed below.
\begin{enumerate}
    \item The energy scale of the domain walls is set by $V\sim \frac{\lambda}{4}v^4 \sim  \frac{9 \beta^2 \xi^2 \Omega_m^2}{2 a_*^6} H_0^2 v^4_\chi(a)
    $, and the energy scale is $\sim 100$ times smaller in model \rom{4} than in model \rom{1}, whereas the above equation predicts $\sim 5 \%$. We defined $v_\chi=\sqrt{1+T_m/\rho_*}$. However, it is complicated because $v$ should be evaluated locally in models \rom{2}-\rom{4}.
    \item The amplitude of the isotropic gradient is $\sim 1$ for both normalisations, indicating that the analytic homogeneous background solution correctly sets the scale in model \rom{1}, whereas the Hubble scale is a good indication in model \rom{4}. The animations show that the gradients {gradually} grow larger with time in models \rom{2}-\rom{5}, while they are the largest for the domain formation and collapse in model \rom{1}.
    \item The equation-of-state parameter $\omega_\phi$ correlates with the domain walls and with the structure, where it is $\omega_\phi\sim 0$. On the other hand, in the underdense parts, it goes towards $\omega_\phi\sim -1$. This agrees with the observation that was made in \cite{christiansen_asevolution_2023}.
\end{enumerate}
We discuss some specific measurements/quantities in more detail below. We start with the matter power spectra.


\subsection{Power spectra}

\begin{figure}[ht]
    \centering
    \includegraphics[width=\linewidth]{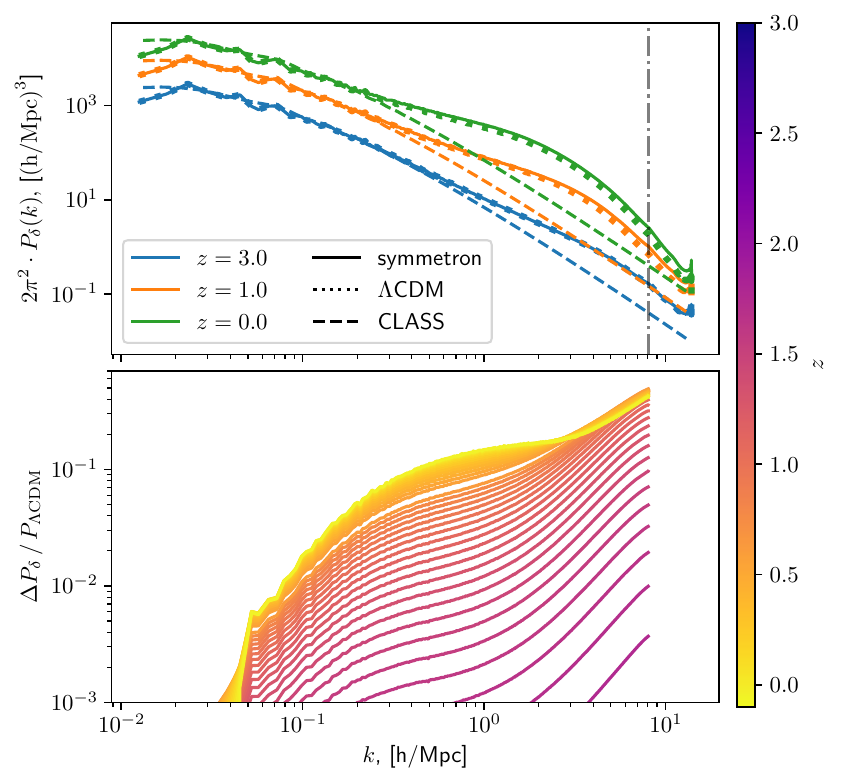}
    \caption{Matter power spectra for model \rom{1} as a function of wavenumber $k$ indicated for different redshifts $z$. Top: Overlaid power spectra from the symmetron asevolution simulation plotted as a solid line, and the corresponding $\Lambda$CDM simulation is shown as dots. The CLASS power spectrum for the same $\Lambda$CDM parameters is shown as a dashed line. Bottom: Relative difference with respect to $\Lambda$CDM of the power spectra from redshift $z=3$ to $z=0$ indicated in graded colours from blue to yellow. The gap in the spectra in the lower panel is not a physical feature and is caused by a typo in the settings file of the simulation. This caused two spectra to be skipped.}
    \label{fig:pk_first_sim}
\end{figure}

\begin{SCfigure*}
    \centering
    \includegraphics[width=1.5\linewidth]{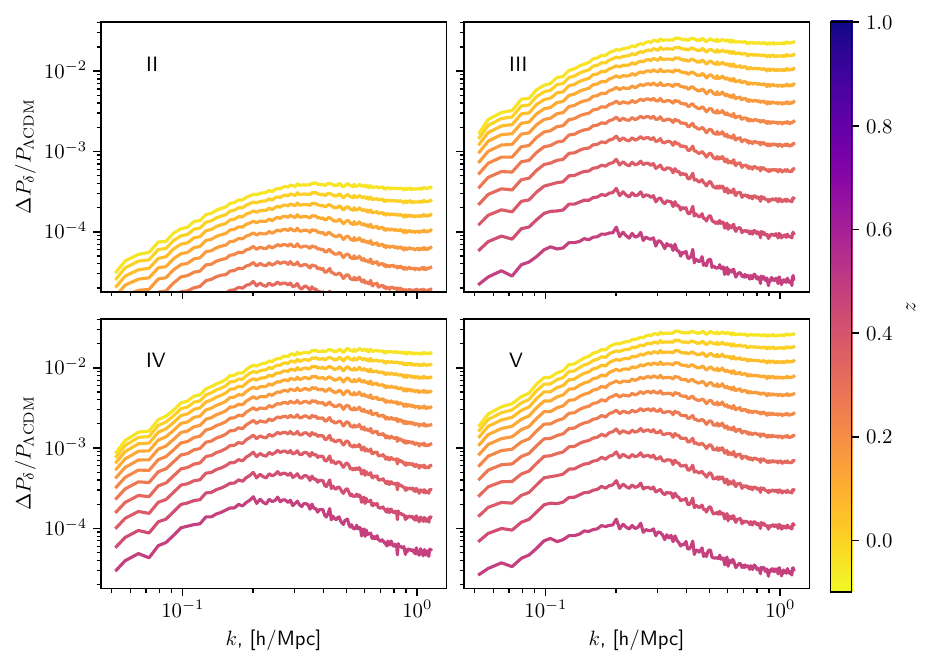}
    \caption{
    \protect\rule{0ex}{5ex}
    Relative difference of matter power spectra with respect to the $\Lambda$CDM case, with redshift $z=1$ to $z=0$ indicated in graded colours from blue to yellow. The spectra are plotted as a function of wave number $k$. The top left to bottom right panels show models \rom{2}-\rom{5}. }
    \label{fig:pk_comparison}
\end{SCfigure*}

In figures \ref{fig:pk_first_sim} and \ref{fig:pk_comparison}, we show the matter power spectrum enhancement relative to the $\Lambda$CDM of models \rom{1} and \rom{2}-\rom{5}, respectively. Figure \ref{fig:pk_first_sim} also shows a comparison of the power spectra to those found by the linear prediction from CLASS \citep{lesgourgues_cosmic_2011}. The agreement is good in general, except at the very largest and smallest scales. This is due to cosmic variance and to the grid resolution, respectively. We expect the power spectra to be reliable to about $k_{\mrm{max}}\sim k_{\mrm{Nyq}}/7$, where $k_{\mrm{Nyq}}=\pi/\mrm d x$ is the Nyquist frequency.
On account of the power spectra, we make the following observations:
\begin{enumerate}
    \item Each of the relative enhancements in the spectra has a bump at a characteristic length scale; this scale seems to be related to the Compton wavelength $L_C$ of the field, and it seems to move towards smaller scales when comparing model \rom{3} and \rom{4}.
    \item The matter power spectrum appears not to be particularly affected by the asymmetry of the potential in model \rom{5} when comparing it to model \rom{3}.
    \item In models \rom{2} and \rom{3}, which have relative force strengths $\beta_{\mrm{\rom{2}}}=1,\beta_{\mrm{\rom{3}}}=8$, respectively, the shapes in the enhancement are similar, but the overall amplitude difference corresponds to $\left(\bar\beta_{\mrm{\rom{2}}}/\bar\beta_{\mrm{\rom{3}}}\right)^2$, as expected from \cite{brax_systematic_2012}, for example.
    \item Finally, comparing models \rom{1} and \rom{2}, we see the effect of varying the critical density $\rho_*\sim a^{-3}_*$. In this instance, shape and amplitude both appear to be affected. The amplitude difference of the bumps is roughly equal to $\left(a_*^{\rom{1}}/a_*^{\rom{2}}\right){^5}$ and approaches $\left(a_*^{\rom{1}}/a_*^{\rom{2}}\right)^4$ for large $k$. Additionally, the power spectrum is less suppressed for larger modes $k$ in model \rom{1}, indicating a lower degree of screening, as expected from the higher critical energy $\rho_*\propto a_*^{-3}$. We expect a different scaling for large $a_*$, and this needs to be investigated separately.
\end{enumerate}
In other words, the results agree with our physical intuition regarding the effect of the parameter choices and with the results obtained in previous approximate treatments of the symmetron in \cite{brax_systematic_2012,llinares_isis_2014}, for instance.

\subsection{Halo mass functions}
\begin{SCfigure*}
    \centering
    \includegraphics[width=1.3\linewidth]{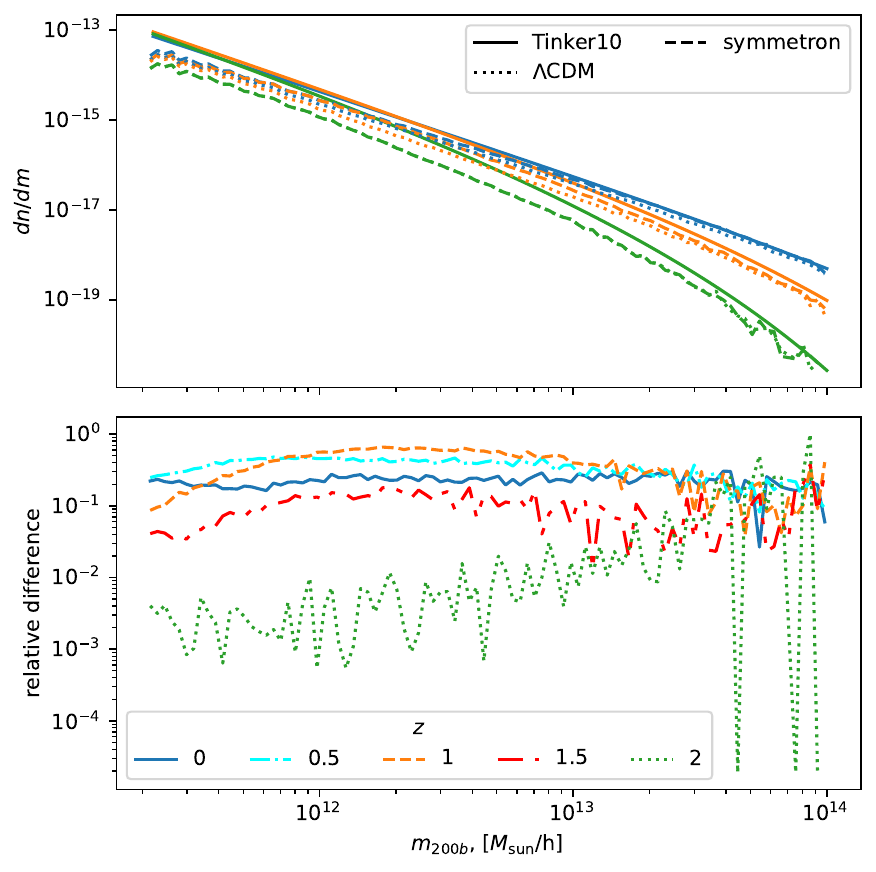}
    \caption{
    \protect\rule{0ex}{5ex}
    {HMFs as a function of halo mass $m_{\mathrm{200b}}$ in units of solar masses $M_{\mrm{sun}}/h$ of the different simulations.} Top: Model \rom{1} HMFs shown for three different redshifts $z=2,1,0$. The HMFs are plotted as dashed lines, compared with the analytic Tinker10 model for $\Lambda$CDM using the CLASS spectra as a solid line, and the result of the $\Lambda$CDM simulation as a dotted lines. Bottom: Relative differences of the HMFs compared to their $\Lambda$CDM counterparts.}
    \label{fig:HMF_model1}
\end{SCfigure*}

\begin{figure}
    \centering
    \includegraphics[width=\linewidth]{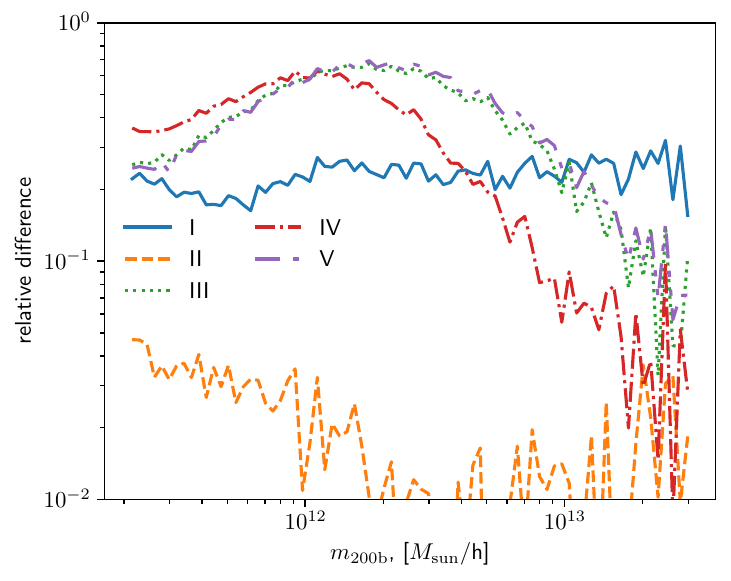}
    \caption{Relative difference between the HMFs of simulations \rom{1}-\rom{5} and $\Lambda$CDM at redshift $z=0$, plotted as a function of the halo mass $m_{\mathrm{200b}}$, in units of solar masses $M_{\mrm{sun}}$. 
    }
    \label{fig:HMF_z0}
\end{figure}

The halo mass functions (HMFs) for models \rom{1} and \rom{2}-\rom{5} are displayed in figures \ref{fig:HMF_model1} and \ref{fig:HMF_z0}, respectively. They indicate the abundances of dark matter haloes at different mass ranges, and are thought to roughly correspond to the galaxy abundances \citep{guo_how_2010}. We found the dark matter haloes using the Rockstar phase-space halo finder \citep{behroozi_rockstar_2013}, and computed the Tinker10 \citep{tinker_toward_2008} $\Lambda$CDM expectation for the HMFs using Pylians \citep{villaescusa-navarro_pylians_2018} and CLASS \citep{lesgourgues_cosmic_2011}. We comment on figures \ref{fig:HMF_model1} and \ref{fig:HMF_z0} below. 
\begin{enumerate}
    \item The top panel of figure \ref{fig:HMF_model1} shows the agreement with analytic estimates from \cite{tinker_toward_2008} constructed out of CLASS power spectra for the $\Lambda$CDM model. The agreement is better for later times $z\rightarrow 0$ and higher masses. This is expected because higher-mass haloes are resolved better at later times, and in addition, more haloes provide improved statistics. We expect that the agreement would be better for all times for larger boxes and better mass resolutions.
    \item The relative differences in the HMFs for model \rom{1} are shown for five different redshifts in the lower panel of figure \ref{fig:HMF_model1}. A general enhancement of $\sim 10 \%$ evolves to $\sim 100\%$ at redshift $z=1$.  That the enhancement appears larger at the intermediate redshift $z=1$ than at $z=0$ is a surprising feature of the derived HMFs. Compared with the relative matter power spectra in figure \ref{fig:pk_first_sim}, this feature is not clearly visible. A small suppression at large $k$-scales at later time can be detected. It might be speculated whether the suppression in the spectra and the HMFs can both relate to the domain collapse at redshift $z\sim 0.5$ by adding kinetic energy to dark matter particles and disturbing screening and bound structures. The enhancement already begins to shrink before the collapse, so that this might not be the best explanation. Another explanation may be that the initial period of enhanced clustering leaves underdensities in the near neighbourhood of the filaments with respect to the $\Lambda$CDM; this would throttle the structure growth in the continuing period. This observation can be made from the density fields in figure \ref{fig:planar_structures_all}, which we present in section \ref{SS:planarstructures}.  The bump in the enhancement in figure \ref{fig:HMF_model1} may be related to the similar enhancement scale viewed in the power spectra of figure \ref{fig:pk_first_sim}
    \item In figure \ref{fig:HMF_z0}, the relative enhancement of the HMFs compared to $\Lambda$CDM for models \rom{1}-\rom{5} is shown at redshift $z=0$. A similar overall amplitude to model \rom{1} is visible in figure \ref{fig:HMF_model1} for models \rom{3}-\rom{5}, 
    indicating that the enhancement might scale with $\Delta A \sim \frac12(\frac{v}{M})^2$, which was kept approximately constant between the two initialisation points. That is, with the exception of model \rom{2}, for which $\Delta A$ is much smaller, and the HMF is suppressed by a factor of $\sim \beta^2$. This is similar to the case with the power spectra in figure \ref{fig:pk_comparison}.
    \item Similarly to the matter power spectra, asymmetricity in the potential, for model \rom{5}, seems to have very little effect on the HMF. A reduced Compton wavelength moved the enhancement towards smaller scales and lower masses, which reduced the level of enhancement of larger haloes. The reduced enhancement of higher-mass haloes would correspond to a more efficient screening mechanism.
\end{enumerate}

\subsection{Oscillating fifth force}
\begin{figure*}[ht]
    \centering
    \includegraphics[width=\linewidth]{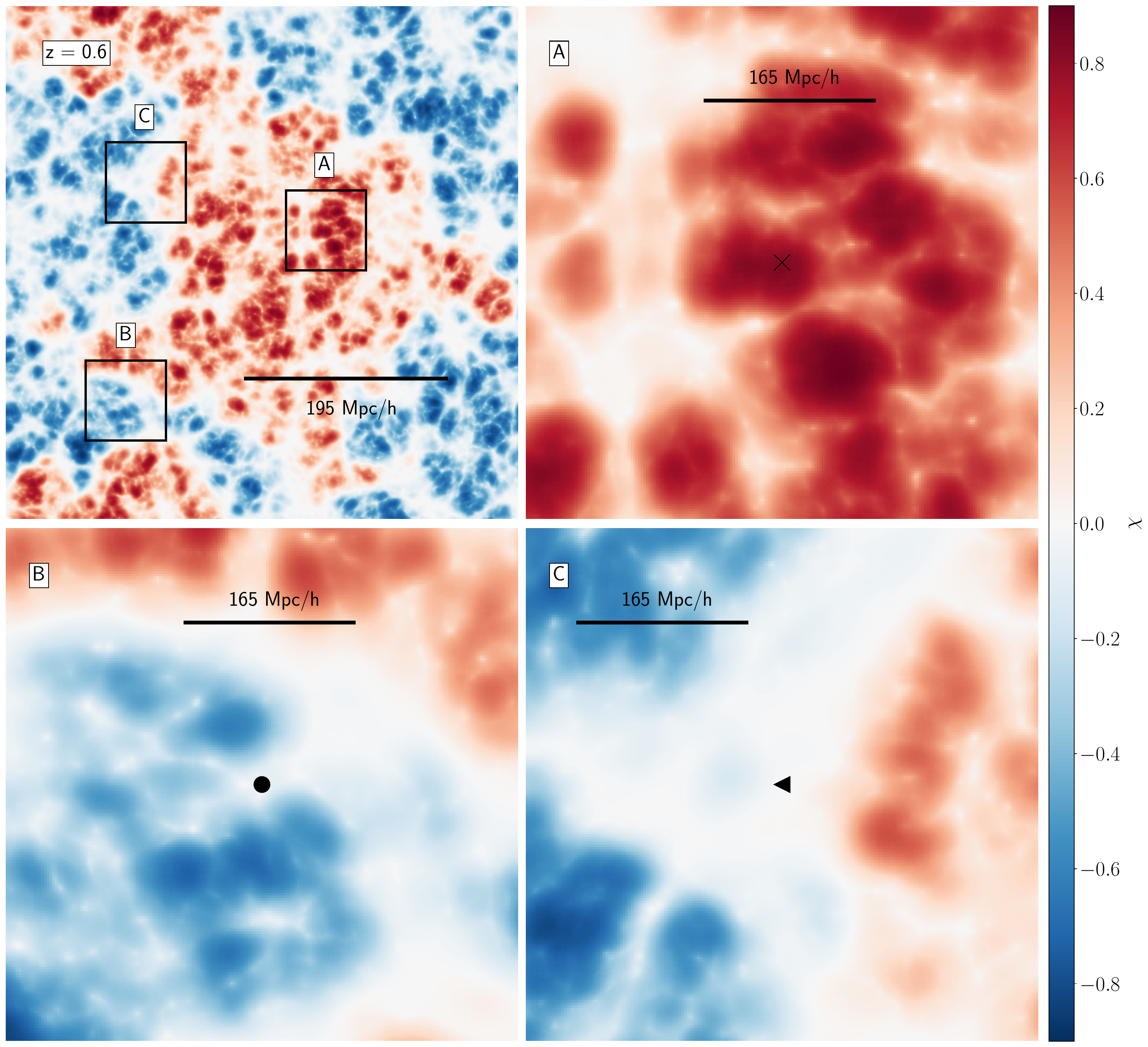}
    \caption{{Different slices of the simulation volume of model \rom{3} at redshift $z=0.6$.} Top {left}: Snapshot of the scalar field. The black squares labelled A-C indicate the zoomed snapshots, which are displayed from the top right to the bottom right panels. The centre mark indicates the point from which the field value was sampled in figure \ref{fig:grav_oscillation}. }
    \label{fig:zoomfigure}
\end{figure*}

\begin{SCfigure*}
    \centering
    \includegraphics[width=1.2\linewidth]{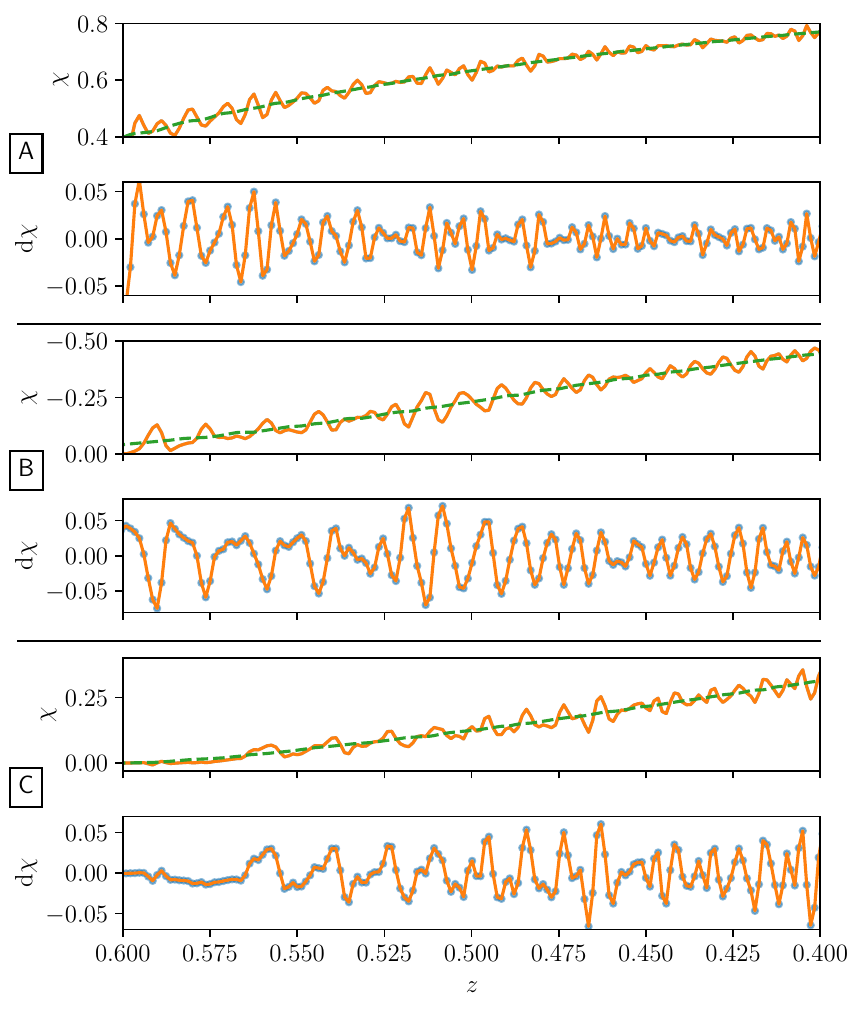}
    \caption{
    \protect\rule{0ex}{5ex}
    Field value $\chi$ in model \rom{3} sampled from a single lattice point as a function of redshift $z$ for the sites labelled A-C, shown in figure \ref{fig:zoomfigure}. The rows {labelled $\rm d \chi$} are the instantaneous field value $\chi$ subtracted the average of the field over the interval $\pm \mrm d\tau/2 =\pm 0.016$ Mpc/h.}
    \label{fig:grav_oscillation}
\end{SCfigure*}

\begin{SCfigure*}
    \centering
    \includegraphics[width=1.5\linewidth]{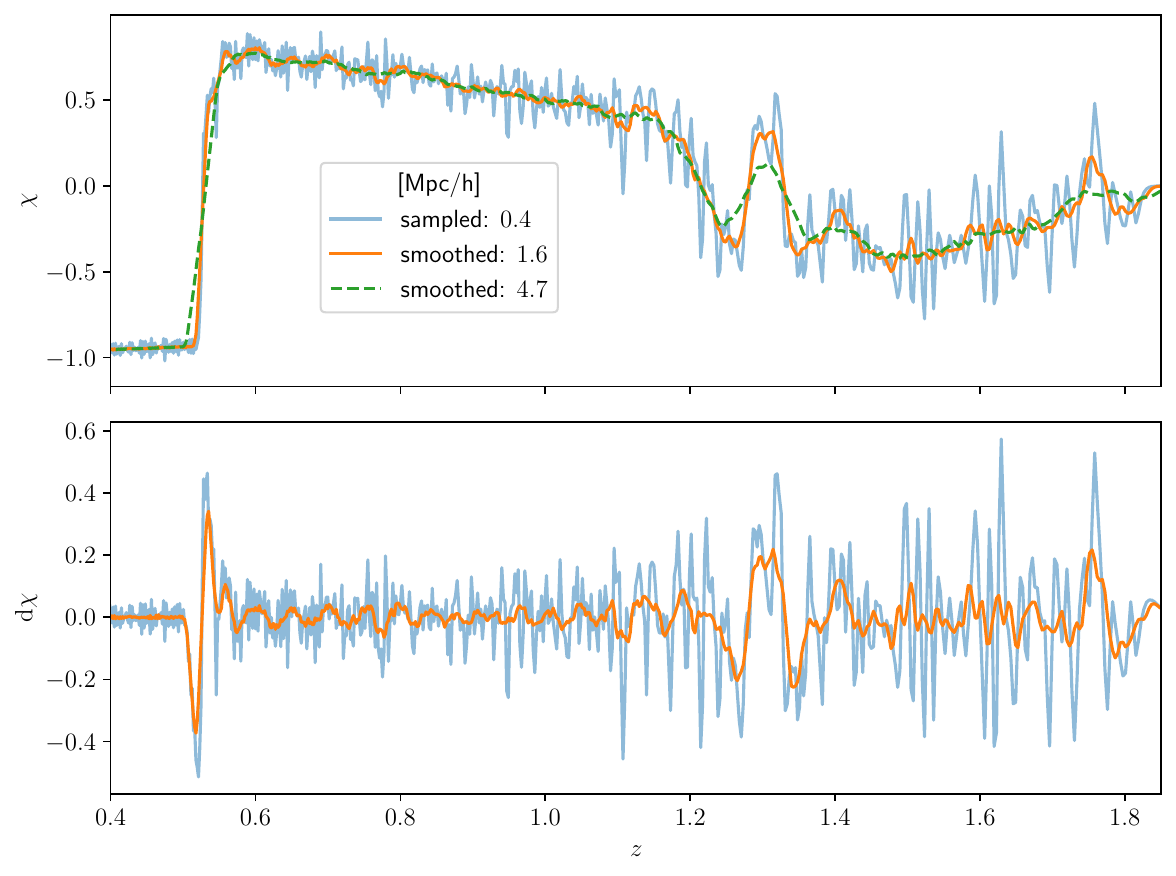}
    \caption{
    \protect\rule{0ex}{5ex}
    Top panel: Field value $\chi$ sampled from a single lattice point as a function of redshift $z$ in model \rom{1}. $\chi$ is the instantaneous field value $\chi$ averaged over the interval $\pm \mrm d\tau/2$, where $\mrm d\tau$ is either $1.6$ or $4.7$ Mpc/h, as indicated in the legend. Bottom panel: Difference between the field values and the field smoothed over $4.7$ Mpc/h. Blue shows the difference for the field value sampled at a $0.4$ Mpc/h rate, and orange shows the difference for the field value averaged over $\pm 1.6$ Mpc/h. }
    \label{fig:oscillating_smoothed}
\end{SCfigure*}

In figures \ref{fig:zoomfigure} from model \rom{3} $(L_C =1$ Mpc/h $,\, z_* =0.1, \, \beta_+ = 8, \, \beta_- = 8)$, the panels labelled A-C show zoomed-in snapshots corresponding to the regions marked A-C in the top left panel. At the centre location of each zoomed-in panel, marked by a cross, circle, and triangle, we sampled the scalar field value throughout the simulation runtime. Part of the resulting time series of the scalar field values is shown in figure \ref{fig:grav_oscillation}. In figure \ref{fig:oscillating_smoothed} we show a similar time series of a randomly selected location in model \rom{1}. We list our observations below.
\begin{enumerate}
    \item All of the locations sampled in figure \ref{fig:grav_oscillation} show a smooth low-frequency evolution, while figure \ref{fig:oscillating_smoothed} shows an extreme low-frequency evolution with several switches in the minimum and additional large-amplitude high-frequency oscillation modes.
    \item Although sampling locations B and C in figure \ref{fig:zoomfigure} both seem to reside in screened regions at redshift $z=0.6$, it appears that neither remains screened. The magnitudes of their background values increase steadily, but the oscillation amplitudes are comparable and more or less constant in time. 
    \item Site C in figure \ref{fig:zoomfigure} shows an effective screening initially, with variations in the field value of $\lesssim 1 \%$. 
\end{enumerate}


\subsection{Planar structures}\label{SS:planarstructures}

\begin{figure*}[ht]
    \centering
    \includegraphics[width=\linewidth]{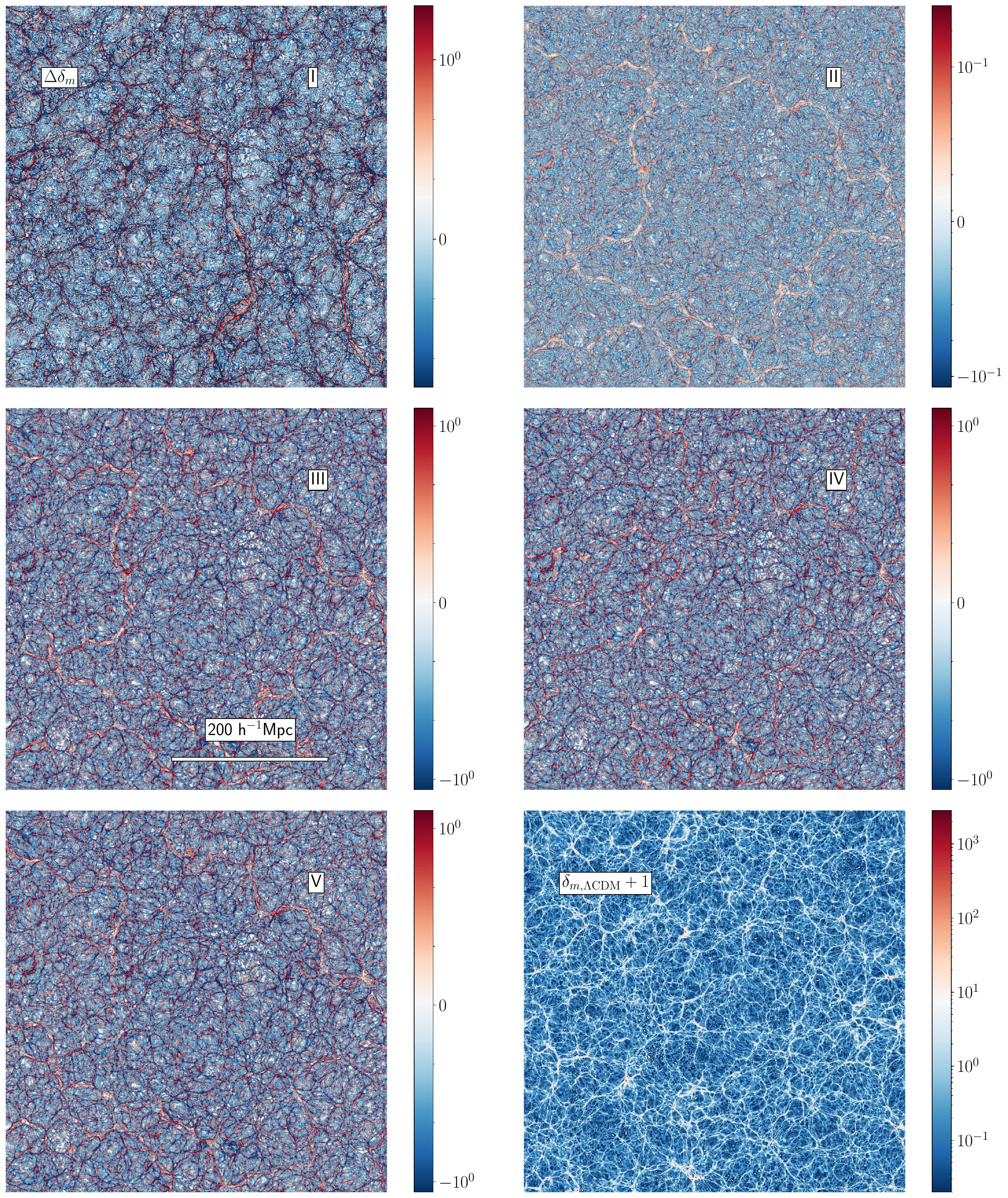}
    \caption{Differences in the overdensity fields of the symmetron models with respect to $\Lambda$CDM, $\Delta\delta_m\equiv \delta_m^{\mrm{symmetron}}-\delta_m^{\Lambda\mrm{CDM}}$. The top left to bottom left panels show models \rom{1}-\rom{5}. The bottom right panel shows the matter density field $\rho_m/\bar\rho_m=\delta_m+1$ for $\Lambda$CDM, where $\bar \rho_m$ is the background value of the density.}
\label{fig:planar_structures_all}
\end{figure*}

In figure \ref{fig:planar_structures_all}, we show the formation of planar structures on cosmological scales by plotting the difference in the overdensity field between simulations of the (a)symmetron and $\Lambda$CDM models. To indicate the structures more clearly, we saturated the colour scale and display it in a symmetric logarithmic plot. We comment on this figure below.
\begin{enumerate}
    \item The largest differences in the overdensities are found in the cosmic filaments for each simulation, where the clustering is stronger and the colour scale is mostly saturated.
    \item The correspondence between the location of the planar structures and the domain walls in figures \ref{fig:4plot} and \ref{fig:field_cartoon} is strong for models \rom{1} and \rom{2}-\rom{5}, respectively.
    \item In the bottom right panel, we show the density field of matter in the $\Lambda$CDM simulation $\rho_m/\bar\rho_m=\delta_m+1$ for comparison. Several of the filaments that make up the skeleton of the domain walls in the (a)symmetron simulation are visible, which may indicate that domain wall pinning has {a} strong effect on the field configuration.
\end{enumerate}
We discuss some more consequences of the planar overdensities in section \ref{S:discussionsandconclusion} and comment on relevant observational signatures.

\subsection{Light propagation}
\begin{SCfigure*}
    \centering
    \includegraphics[width=1.5\linewidth]{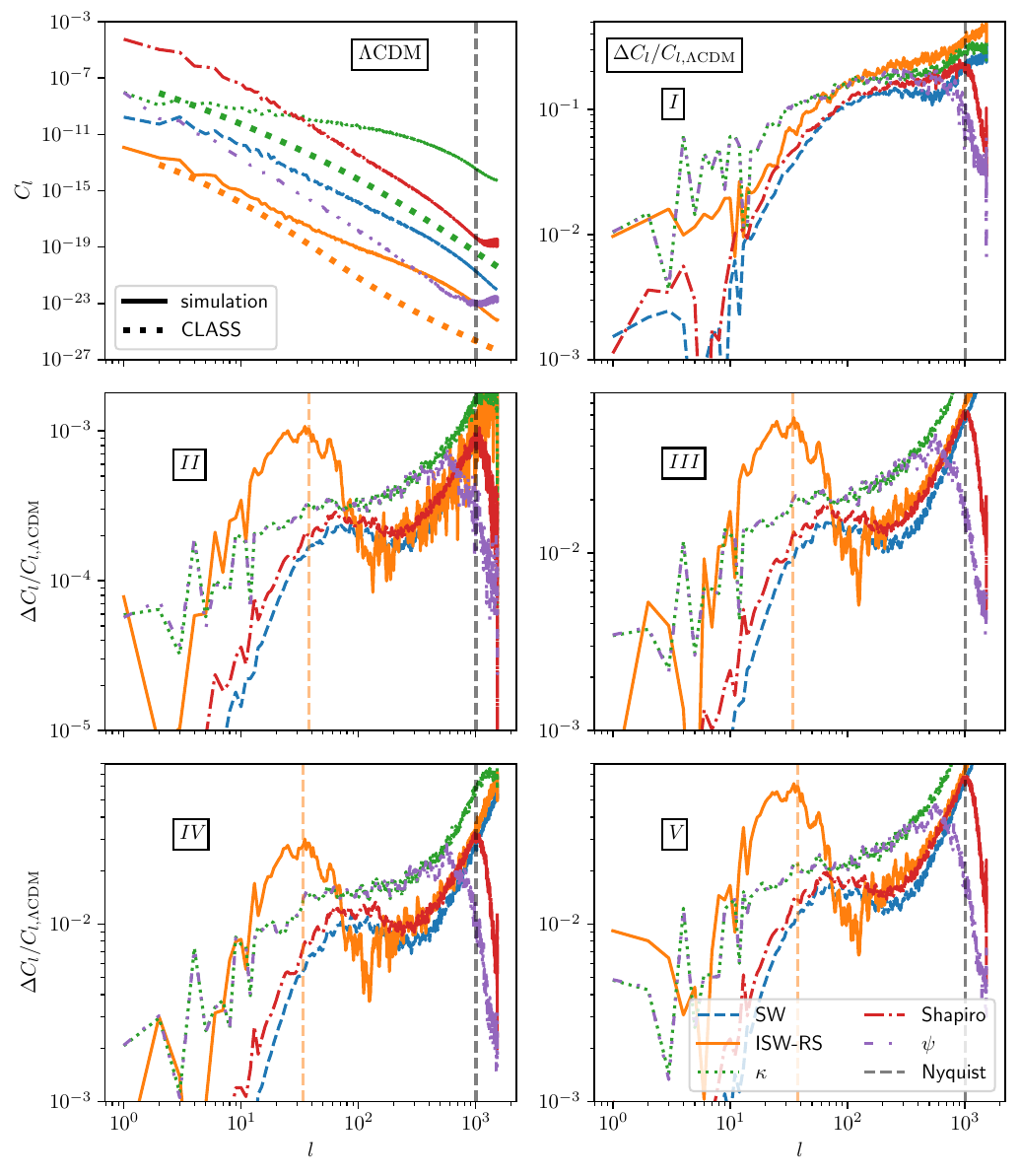}
    \caption{
            \protect\rule{0ex}{5ex}
    Secondary light propagation effects in the full-sky light cone of an observer at redshift $z=0$ located at the origin position $\vec x_0=(0,0,0)$ Mpc/h. The source field is located at redshift $z=0.0835$. Top left: Angular power spectra of the $\Lambda$CDM simulation shown as thin lines. The thick dotted line shows the CLASS estimate, which is only available for the ISW, convergence, and lensing potential. Top right to bottom right: Relative angular power spectra with respect to $\Lambda$CDM for the simulations of models \rom{1}-\rom{5}. The meaning of the labels in the legend is SW: Ordinary Sachs-Wolfe effect. ISW-RS: Integrated Sachs-Wolfe and non-linear Rees-Sciama effect. $\kappa$: Convergence field. Shapiro: Shapiro time delay. $\psi$: Lensing potential.}
    \label{fig:lightcone_effects}
\end{SCfigure*}

\begin{SCfigure*}
    \centering
    \includegraphics[width=1.5\linewidth]{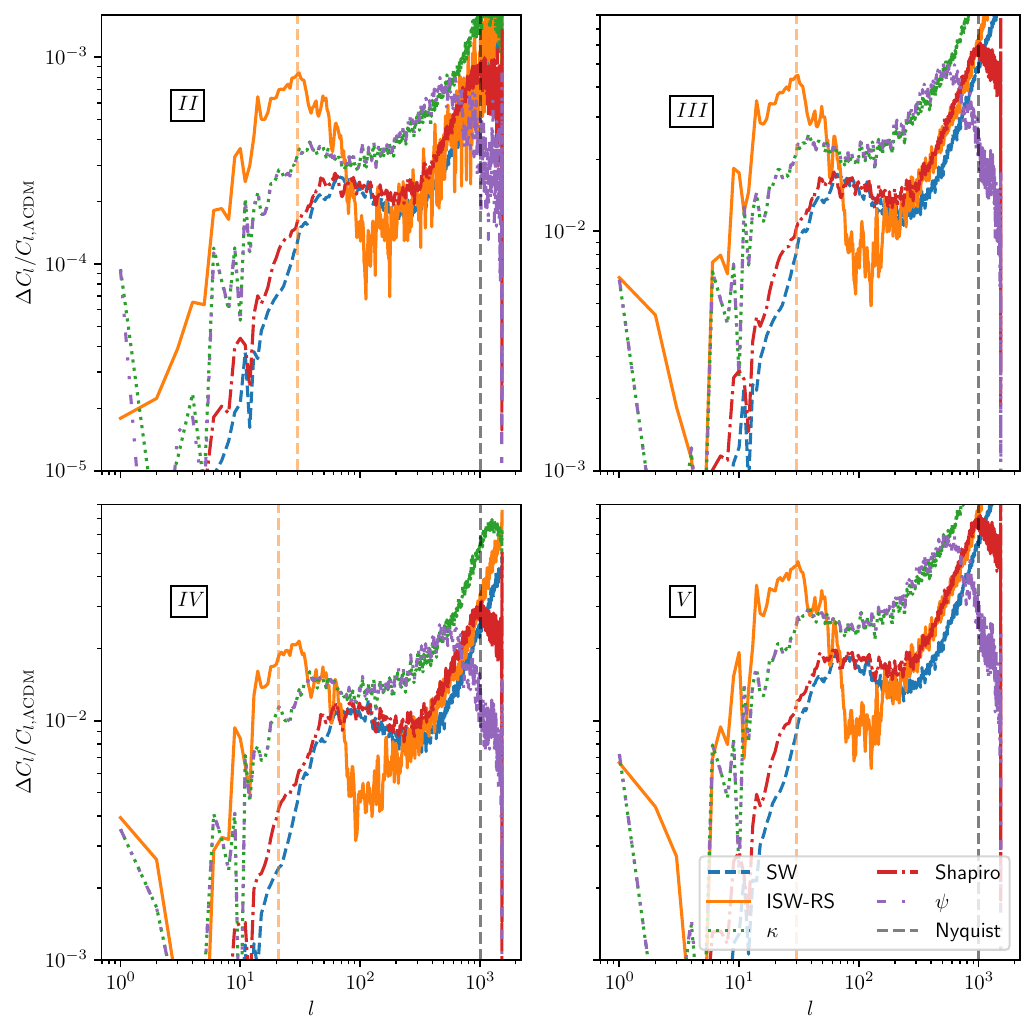}
    \caption{
            \protect\rule{0ex}{5ex}
    Secondary light propagation effects in the full-sky light cone of an observer at redshift $z=0$ located at the centre position $\vec x_0=(250,250,250)$ Mpc/h. The source field is located at redshift $z=0.0835$. Top left to bottom right: Relative angular power spectra with respect to $\Lambda$CDM for the simulations of models \rom{2}-\rom{5}. The meaning of the labels in the legend is SW: Ordinary Sachs-Wolfe effect. ISW-RS: Integrated Sachs-Wolfe and non-linear Rees-Sciama effect. $\kappa$: Convergence field. Shapiro: Shapiro time delay. $\psi$: Lensing potential.}
    \label{fig:lightcone_effects2}
\end{SCfigure*}

We expect light to be redshifted by the expansion of space. Additionally, secondary effects arise from the interaction of the light with inhomogeneities that will perturb its redshift and direction.
These secondary effects on light from a source field at $z=0.0835$ are shown in figures \ref{fig:lightcone_effects} and \ref{fig:lightcone_effects2}. The source field is limited to a distance $d(\mrm{source})\leq B/2=250$ Mpc/h, where $B$ is the box size of the simulation, corresponding to the chosen source redshift. The light-cone was constructed within gevolution on the fly, and the integration of the null geodesics from the source field was made by a shooting method that was presented in \cite{adamek_bias_2019}. In figure \ref{fig:lightcone_effects} we show the angular power spectrum of the secondary effects due to the {(ISW-RS)} effect, lensing {potential}, convergence ($\kappa$), Shapiro time delay, and the gravitational redshift (or the ordinary Sachs-Wolfe effect (SW)). These effects{,} their computation and observation were discussed in \cite{hassani_clustering_2020}. Because the distance to the source, $r< 250$ Mpc/h, is small, the features of the angular power spectra correspond to physical scales, $d$, which are small, approximately
 \begin{align} \label{eq:lightconescale}
    l\simeq \pi/\Delta\phi \simeq r\pi/d ,
 \end{align} 
for $\Delta\phi\ll 1$, where $l$ is the moment of the spherical harmonic corresponding to the azimuth. We first consider the light cone of an observer at redshift $z=0$, located at the edge of the simulation domain $\vec x = (0,0,0)$ Mpc/h (figure \ref{fig:lightcone_effects}): In all of the late-time symmetry-breaking simulations, models \rom{2}-\rom{5}, the signature in the ISW signal is clear, giving a $\sim 10\, \%$ relative enhancement with respect to $\Lambda$CDM and peaking at a characteristic scale of $l\sim 20-40$. The signal is suppressed in model \rom{2}, where $\beta_*$ is smaller, but has an identical shape. The remaining secondary effects  in the first light cone share a similar shape and peak around $l\sim 100$, although the convergence field $\kappa$ is slightly larger and peaks at larger scales $l\sim 40-100$. We find the same qualitative behaviour in the second light cone of an observer at redshift $z=0$ at the centre of the box $\vec x_0 = (250,250,250)$ Mpc/h, although the convergence of the angular power spectrum is larger. It peaks more sharply at $l\sim 40-50$. The secondary effects on the light propagation in model \rom{1}, for which we produced the first light cone alone, have a very different appearance with far weaker peaked angular spectra. The enhancements on the ISW-RS and convergence are still largest and reach $\sim 20\%$ at small angular scales $l\sim 200$.
We place these observations into context from the literature and discuss some observational signatures in the discussions section 
\ref{S:discussionsandconclusion}.

\section{Discussions and conclusion} \label{S:discussionsandconclusion}

In this section, we discuss some of the observations made in the simulation data in section \ref{S:hrsims} and their implications for the features of the models and signatures in observational surveys.

\subsection{Observations made in simulation data} \label{SS:dataobservations}
For domain partitioning and domain wall stability, considering the simulations listed in table \ref{tab:modelparameters} and displayed in figure \ref{fig:background_overview}, and those shown in appendix \ref{A:parameters} in figures \ref{fig:A:parameter_stability} and \ref{fig:A:parameter_collapses}, we list our observations below.
\begin{enumerate}
    \item The domain wall stability seems to be improved when $L_C$ and $z_*$ are decreased.
    \item The domain wall stability and partitioning seems to be almost unaffected when $\bar \beta$ is varied. However, for larger $\bar\beta$s, we expect the secondary effect of enhanced clustering along the walls to which in turn the domain walls are pinned, which enhances stability.
    \item Although we were unable to find stable domain walls for asymmetric potentials $\Delta\beta\neq 0$, we were able to make them quasi-stable on presumably arbitrarily long timescales by sufficiently decreasing $L_C$ and $z_*$. 
    \item The domain wall instability and the effect on partitioning caused by the asymmetric potential seem to depend only on the relative level of asymmetry $\Delta\beta/\bar \beta$. {The stability} was not affected when this was kept constant, but the $\beta$s were changed.
    \item By decreasing the Compton wavelength, we find less variability in the fraction of positive minimum, meaning that the domain size has decreased so that we resolved more minima. This means that the fraction of positive minima is $\sim 0.5$, as expected from the initial $\mathbb Z_2$ symmetry.
\end{enumerate}
These observations suggest that for this part of the parameter space, the pinning of domain walls to the filaments of large-scale structure is enhanced by the higher density contrasts in the filaments (later-time symmetry breaking) and by the interaction of the field with smaller-scale structures (smaller $L_C$). Intuitively, we would expect there to be a feedback process in which the enhanced clustering in the domain walls by itself would increase their stability by either increasing the filament density contrasts or creating wall structures where no filaments were originally. We see no strong indications that this were the case, but this effect might become more important at higher values of $\beta$. Some {subtle} differences between model \rom{2} and \rom{3} are visible in figures \ref{fig:field_cartoon} and \ref{fig:aq_cartoon}, such as the {$z=0.2$} collapse of the small centre-left positive-minimum island {(approximately located at coordinates $(0.1,0.4)B$, where $B$ is the box size)}, indicating that this effect is not entirely absent.

For the explored late SSB parameter space of the symmetron, the mismatch
n all instances of $z_*\lesssim 0.7$ is increasingly large between the selected time of symmetry breaking at the background $z_*$ and the actual time $z_{\mrm{SSB}}$, $\Delta z=z_{\mrm{SSB}}-z_*$ for increasingly smaller $z_*$. This is very interesting because it means that $0 \lesssim z_{\mrm{SSB}}\lesssim 1$ 
for a large part of the parameter space $(z_*,L_C)$. When this part of the symmetron parameter space were linked to the dark energy phenomenon of late-time cosmological acceleration, the coincidence problem would be explained because $z\lesssim 1$ marks the onset of high density contrasts. We will explore this possibility in more detail in a future work.

\subsection{Signatures}

In the previous section, \ref{S:hrsims}, we listed several results. We discuss their significance below.

The peaked ISW-RS signal in figures \ref{fig:lightcone_effects} and \ref{fig:lightcone_effects2} is especially interesting in the context of the anamolous ISW signals that were claimed to be detected in large voids (see \cite{kovacs_imprint_2017,kovacs_part_2018,kovacs_more_2019,vielzeuf_dark_2020}). Intuitively, we expect voids to be repulsive due to the fifth force $F_5\sim -\partial_r \phi^2$, and we therefore expect an enhanced clustering towards the filaments that enclose voids compared to $\Lambda$CDM. Figure \ref{fig:planar_structures_all} shows that this indeed occurs: the filaments and domain walls all have enhanced densities (red) 
, while regions immediately next to filaments or within voids have lower densities (light and dark blue). The effect on the evolution of the potentials from the different clustering is imprinted on the ISW signal. The important scale for this for models \rom{2}-\rom{5} based on equation \eqref{eq:lightconescale} is $\sim 18$ Mpc/h for the mean distance of $R_{{\mrm{lightcone}}}/\sqrt{3}\sim 144$ Mpc/h. Since this scale does not appear to depend significantly on $\xi_*$ because there is no clear difference of it in model \rom{4}, we speculate that it may be set by the parameter $a_*$ instead, or equivalently, by $\rho_*$. This might be related to the scale on which $\delta_u\bar\rho_m<\rho_*$, where $\delta_u$ is the underdensity of voids. This should be examined quantitatively. The width of the peak of the ISW signal instead seems to depend on the Compton wavelength and possibly also on $a_*$, although this needs to be verified in detail. In a future work, we will investigate the parameter dependence of this signal and study the ISW signal of voids in particular.

The planar structures that we identified
in figure \ref{fig:planar_structures_all} are interesting in the context of similar features that were claimed to be observed in the local universe (see e.g. \cite{peebles_flat_2023}). It is especially interesting for structures that were claimed to be detected on scales larger than what the standard $\Lambda$CDM should be able to account for, such as the detected dipoles in large-scale structure \cite{secrest_test_2021,secrest_challenge_2022,panwar_probing_2023} or other claimed structures such as the Hercules-Corona Borealis (HCB) Great Wall \citep{horvath_new_2015,horvath_clustering_2020}. The structures that appear in our simulations (figure \ref{fig:planar_structures_all}) extend across the box of $\sim 500$ Mpc/h and remain approximately flat on scales up to $100$-$250$ Mpc/h in some cases. Their overdensity amplitude can be adjusted by a change in $\bar\beta$, and their scale seems surprisingly not to depend strongly on the parameter choices within our explored parameter space.
 The scale in the late-time simulations seems to come from the cosmic filaments and the scale at which they encapsulate domains. We expect variations in this scale for more extreme variations in the parameter space. We leave this study to a future work. The planar scales we found in the simulations are comparable to those reported in \cite{peebles_anomalies_2022}, who reported flat sheet-like structures with a thickness of $\sim 5$ Mpc 
 and lengths of $\sim 150$ Mpc. An interesting future work would be to study the ability of the (a)symmetron parameter space to account for such observations in detail, over a larger parameter space, and using CDM haloes instead of particles. The haloes would correlate more stronger with the galaxy number count statistics made use of in \cite{peebles_flat_2023}. Furthermore, it would be interesting to analyse the full three-dimensional data to learn more about the sheet structure morphologies.

The observation of local fluctations in the scalar field, and thus also the fifth force, $F_5\sim \partial_r(\phi^2)$, in figures \ref{fig:grav_oscillation} and \ref{fig:oscillating_smoothed} may have distinct observational signatures. The effect of oscillating ultralight scalar fields has been considered in the literature for coupling to the dynamics of pulsars \citep{unal_probing_2022} and to the oscillations of stars \citep{sakstein_dark_2023}, which would help to impose strict constraints on the allowed energy density parameter of it when considered as a dark matter candidate. In addition to the effect of an oscillating gravitational potential owing to the energy density of the scalar field, we have the oscillating fifth force in this instance, which might work towards imposing stricter constraints on the model parameter space. However, screening must be taken into account, which was shown in figure \ref{fig:grav_oscillation} to dramatically suppress the oscillation and the overall scalar field value for the duration of the screening activity.
In the past, the possibility of disrupting the screening mechanism from so-called cosmic tsunamis was reported \citep{hagala_cosmic_2017}. These are waves in the scalar field. A study of such disruption events in the simulation output will be made in the future. In addition to pulsars and stars, periodic systems on different length scales may help to constrain dark matter masses and the fifth force couplings that operate on these scales. In our simulations, the oscillation has a timescale of $\sim$ Mpc/h, in which case it might affect the dynamics of galaxy clusters. Likewise, we expect the oscillations in models with $L_C\sim$ kpc/h to be able to affect the dynamics of galaxies. For example, when the attractive potential of a cluster changes, the virial radius oscillates and prevents the particles from having stable orbits; for one instance, the force increases, giving the particles a higher centripetal acceleration and injecting kinetic energy; and in the next moment, the force decreases, and the particles become unbound and spread out. This was observed in \cite{dutta_chowdhury_random_2021} for an oscillating potential, and we expect the effect to be enhanced when a fifth force is included. It would be an interesting project to consider whether this dynamic would be able to account for the velocity dispersion of stars in galaxies for choices of $\sim$ kpc/h scale general Compton wavelengths. When the screening is included, the effects on systems with AU timescales might be considered for shorter Compton wavelengths to constrain lunar ranger experiments \citep{xue_precision_2020}.

Enhanced clustering owing to the fifth force is visualised in the matter power spectra in figures \ref{fig:pk_first_sim} and \ref{fig:pk_comparison} and in the HMFs in figures \ref{fig:HMF_model1} and \ref{fig:HMF_z0} for models \rom{1} and \rom{1}-\rom{5}, respectively. Overall, the effect of the scalar field is to enhance clustering, but this is not uniform over the scales $k$; the relative spectral enhancements in figure \ref{fig:pk_comparison} are clearly peaked and then suppressed for larger and smaller scales. At larger scales, we expect the suppression to be caused by the horizon $H^{-1}$, and on smaller scales by screening $\delta_m(k)\rho_m>\rho_*$. The enhanced clustering presents an obstacle to the model, as the effect is likely to worsen the $\sigma_8$ tension reported in the literature \citep{mccarthy_flamingo_2023}, which would require a stronger mechanism to suppress structure formation in the late-time universe than what is needed for $\Lambda$CDM. The $\sigma_8$ tension is only $\sim 2-3\sigma$ in significance so far, which means that it is not clear whether it will persist as more data are included, however. Observations that are more consistent with enhanced clustering are also indicated in the literature
\citep{mazurenko_simultaneous_2023}. Speculatively, thinking about suppression of clustering, a mechanism on cosmological scales might be expected for a model where the (a)symmetron makes up dark matter and has a conformal coupling so that it has a $\sim$Mpc/h scale Compton wavelength scale in screened regions during the time between the last scattering surface and now. In this instance, it would operate similarly to fuzzy dark matter \citep{niemeyer_small-scale_2020} cosmologically. With growing inhomogeneities at late times, higher-mass dark matter particles might be accounted for and might simultaneously have a Compton wavelength of $\sim$kpc/h in galaxies today. Furthermore, the correlation ranges of dark matter particles in unscreened regions 
will correspond to the choice of the Compton wavelength, which can be chosen at cosmological scales $\sim$Mpc/h. In this instance, suppression and enhancement can both be derived depending on the environment.
These suggestions require a more careful analysis and will be explored in a future work. For the simulations considered here, the energy budget of the symmetron was chosen to be almost negligible, and reliably simulating fuzzy symmetron dark matter would present an additional computational challenge that would be an interesting endeavour to consider. The shrinking relative enhancement observed in the HMF of model \rom{1} presents another opportunity for not worsening the $\sigma_8$ tensions as much. Structure formation may continue into a period of weaker clustering due to a depletion of matter in neighbouring environments.

\subsection{Conclusion}

We have presented the first cosmological simulations of the (a)symmetron for which 
spatial and temporal convergence in the field configuration was demonstrated, and we retrieved quantitative estimates of novel signatures such as secondary light-propagation effects, the formation of sheet-like structures imprinted on the dark matter distribution, 
and local oscillations in the fifth force. We analysed the evolution of the power spectra and the HMFs. In the latter, we observed a feature of greater enhancement at intermediate times, the cause of which will be interesting to determine in the future. The possibility of producing simulations that fully resolve the dynamic part of the symmetron parameter space opens several possible paths towards uncovering new phenomenological features of the model and further constraining its parameter space. In the future, we wish to study the ISW-RS signal in more detail, its correlation with void structures, and in particular, the evolution of the gravitational potential of large voids, taken in the context of the anomalous ISW signal found in the literature \citep{kovacs_imprint_2017}. We wish to study the effect of oscillating potentials on the radial mass profiles of dark matter haloes in greater detail, and we would like to conduct a more extensive study of the imprint of the domain walls on the dark matter density field, where it creates planar structures, taken in the context of \cite{peebles_flat_2023}. In the future, a larger project would be a simulation of the (a)symmetron in the part of its parameter space where it accounts for dark matter, which is well motivated in the literature \citep{burrage_radial_2017,burrage_symmetron_2019,kading_lensing_2023}, in addition to our findings presented here and in the companion paper \cite{christiansen_gravitational_2024}. The (a)symmetron model is an interesting and phenomenologically complex model that includes mechanisms for solving many of the observational tensions of modern day cosmology, along with a large array of new and unique predictions. With the improvement of computational resources and techniques, we expect to access predictions in a larger part of its parameter space in the near future.


\section*{Acknowledgements}
We would like to thank Julian Adamek and Jens Jasche for helpful discussions or comments. ØC would like to thank Constantinos Skordis and the Central European Institute of Cosmology (CEICO) for hosting him during the final stages of the project. FH wishes to thank Martin Kunz and Jean-Pierre Eckmann for their insightful discussions and hosting him during the project's early stages. We thank the Research Council of Norway for their support.  The simulations were performed on resources provided by 
UNINETT Sigma2 -- the National Infrastructure for High Performance Computing and 
Data Storage in Norway. This
work is supported by a grant from the Swiss National Supercomputing Centre
(CSCS) under project ID s1051. 

----------------

\begin{appendix}
\section{Parameter space}\label{A:parameters}

\begin{figure}[ht]
    \centering
    \includegraphics[width=\linewidth]{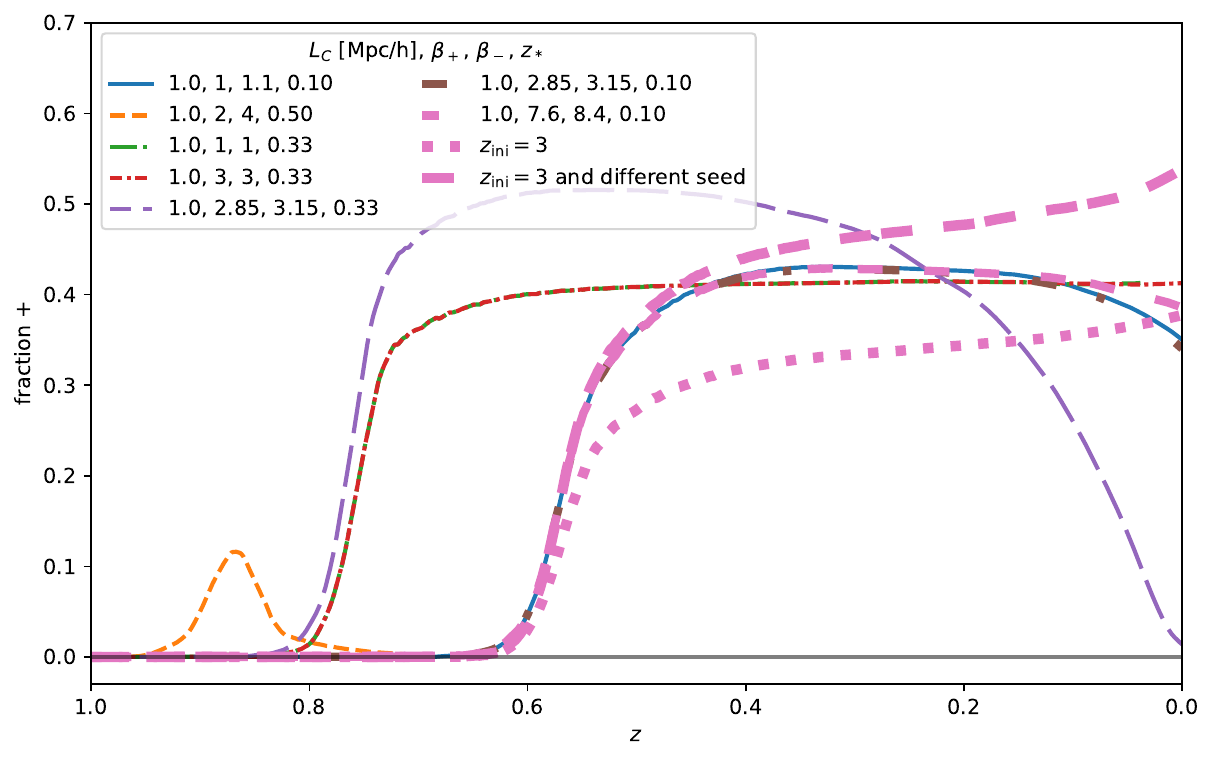}
    \caption{
    Fraction of the simulation volume with the scalar field in the positive minimum as a function of redshift for a range of (a)symmetron parameters. The last three lines in pink all show the same parameter choice, but at an earlier initialisation point as the dotted line, and with a different seed number as the long dash line. }
    \label{fig:A:parameter_stability}
\end{figure}

\begin{figure}[ht]
    \centering
    \includegraphics[width=\linewidth]{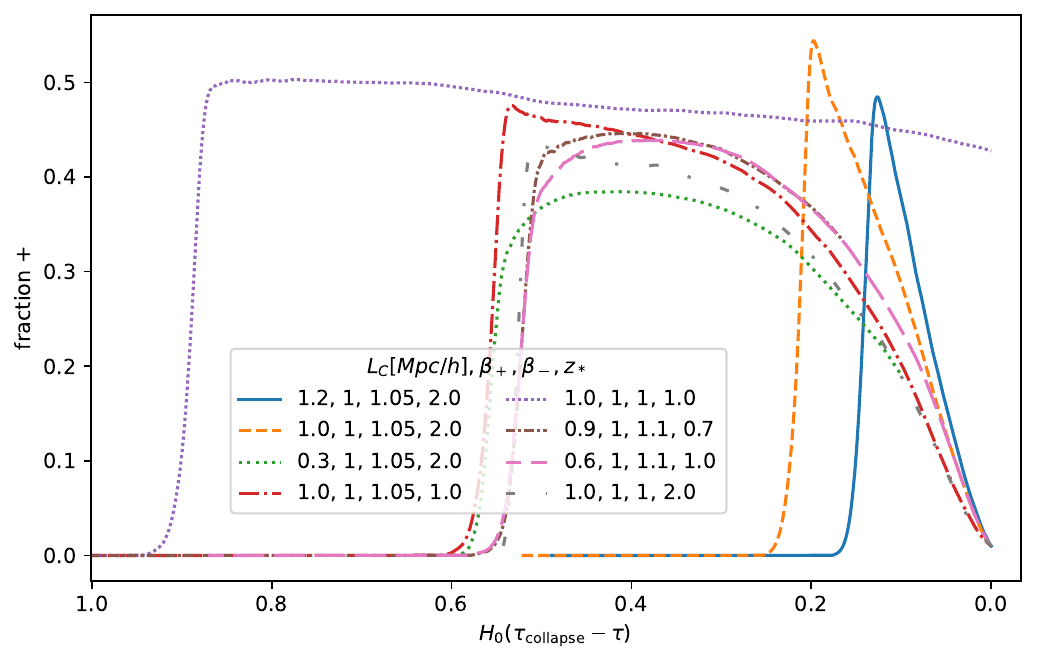}
    \caption{
    Fraction of the simulation volume with the scalar field in the positive minimum as a function of conformal time before collapse (normalised by the Hubble scale) for a range of (a)symmetron parameters. In the case of no collapse (densely dotted purple), it is instead plotted against the conformal time until the end of the simulation $z=0$.}
    \label{fig:A:parameter_collapses}
\end{figure}

In preparation of the more expensive simulations presented in section \ref{S:hrsims}, we ran a large number of $N_{\mrm{grid}}=512^3$, $\left( 500\,\mrm{Mpc/h} \right)^3$ volume simulations 
in order to identify an interesting region in the parameter space that permitted the existence of quasi-stable domain walls throughout the simulations.
As a simple quantity to quantify the stability and behaviour of the domain walls, we compared the fraction of the simulation volume in the least occupied {domain} (labelled + here). The result is shown in figures \ref{fig:A:parameter_stability} and \ref{fig:A:parameter_collapses}. By choosing a resolution that barely resolved the Compton wavelength $\mrm d x = 0.9 \cdot L_C$, but {keeping} the size of the simulation domain $L_{\mrm{box}}=500$ Mpc/h, we expected the results to be representative for the dynamics in the higher-resolution simulations of section \ref{S:hrsims}, according to section \ref{S:convergence} and figure \ref{fig:idealCaseSpatialRes}. Several of the observations listed in section \ref{SS:dataobservations} on the effect of variation of the symmetron parameters on the domain wall stability can be made from the {figures \ref{fig:A:parameter_stability} and \ref{fig:A:parameter_collapses}}.

\FloatBarrier

\section{Initialisation}
As was demonstrated in subsection \ref{SS:timescale}, resolving the Klein-Gordon equation in the pre-symmetry-broken time at early redshifts $z\sim 100$ would be computationally demanding, both because the Compton scale is smaller due to the high mass $m\sim \sqrt{-T}$ and because we would have to evolve the symmetron for a longer duration. In subsection \ref{SS:timescale}, we argued that a late-time initialisation was sufficient and would save computational expenses. We substantiate this argument here. 

\begin{figure}[ht]
    \centering
    \includegraphics[width=\linewidth]{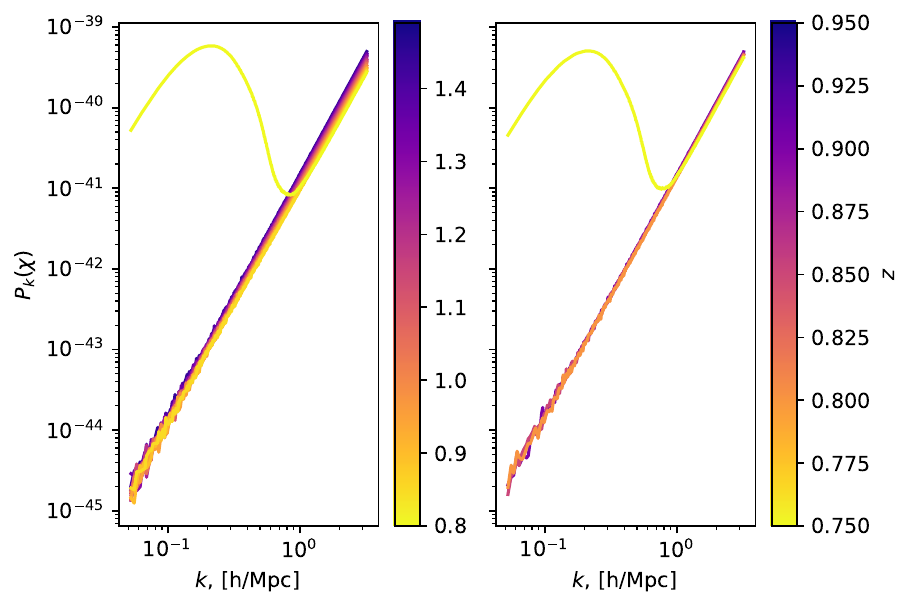}
    \caption{Pre-symmetry-breaking evolution of the scalar power spectrum for a simulation of $512^3$ grids and $(500\,\mrm{Mpc/h})^3$ volume. The left panel has an initialisation time at redshift $z_{\mrm{ini}}=1.5$, and the right side panel has $z{_\mrm{ini}}=1$.}
    \label{fig:ICcheck}
\end{figure}

\begin{figure*}[t]
    \centering
    \includegraphics[width=\linewidth]{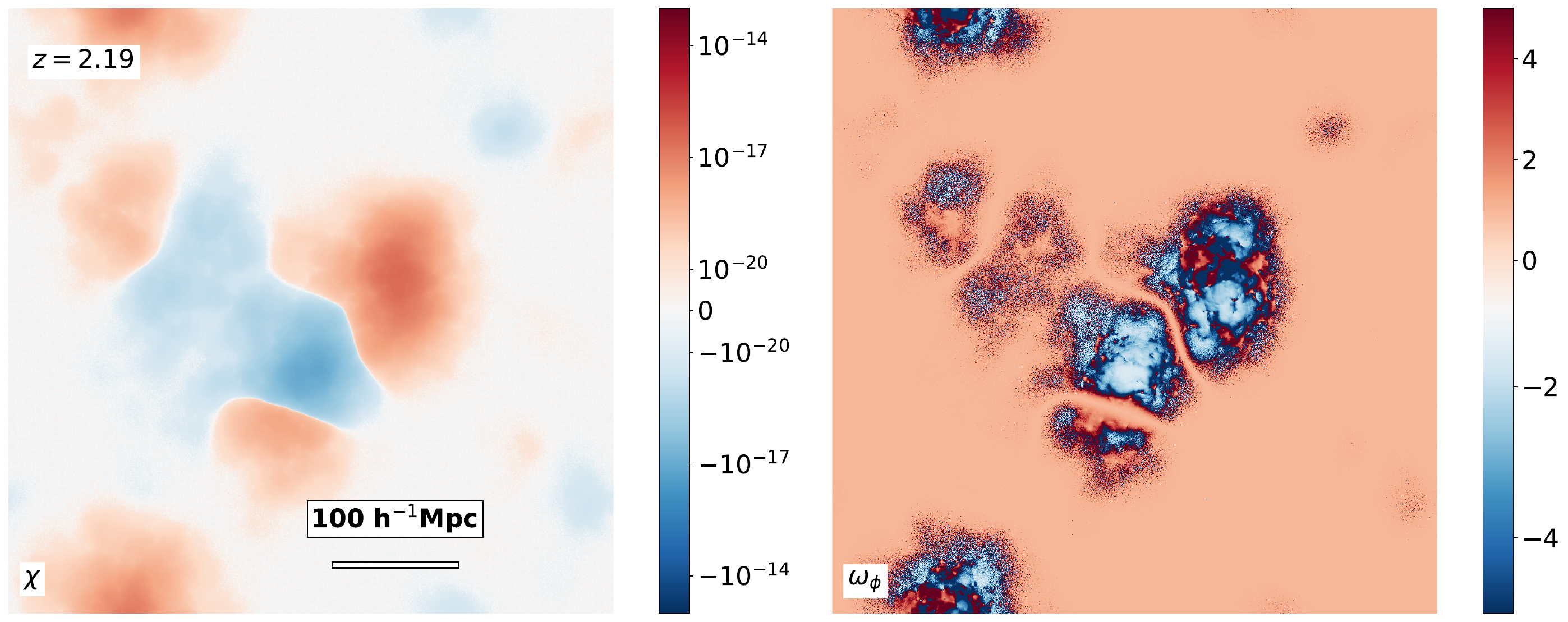}
    \caption{
    Evolution in the scalar field, $\chi$, of model \rom{1} before the global symmetry breaking, which occurs at $z_{\mrm{SSB}}\sim1.9$. The right panel shows the equation-of-state parameter of the field, which evolves until the global symmetry breaking and indicates the final domain walls.}
    \label{fig:prevolution}
\end{figure*}

The initialisation that was used in \cite{christiansen_asevolution_2023} and {that} was also {used}  here{,} initialises the scalar field as a draw from a Gaussian scale-invariant power spectrum of some small amplitude set by the user $A_{\mrm{amp}}\sim 10^{-20}$. The following evolution of the field in the pre-symmetry-broken regime involves local oscillations {on spatial scales} of the Compton scale $L$. The energy scale made use of here for the symmetron is $\sim 10^{-6}\cdot\rho_{c,0}$ at most post-symmetry breaking, and it is completely negligible before $\sim 10^{-40} \cdot \rho_{c,0}$, so that we neglect no relevant gravitational effects. The fifth force is $\propto \phi \partial_r \phi$, and it is therefore also suppressed because $\phi\sim A_{\mrm{amp}}$. The only information that might be lost by dismissing the early time field evolution is then the phase information of the field and the correlation structure that would be established throughout this time. In figure \ref{fig:ICcheck} we demonstrate that the initially scale-invariant power spectrum of the scalar field remains very nearly scale-invariant up until the symmetry-breaking time, and the symmetry-breaking process otherwise takes place identically. Similar features for earlier-time initialisations are observed, although there is an amplitude modulation owing to the Hubble friction acting on the field. Since both the amplitude and initial phase of the scalar field were chosen arbitrarily, the information of its evolution in the pre-symmetry-broken time is not relevant as long as it leaves its statistics qualitatively similar. Small variations in the amplitude seems to affect the symmetry-breaking process only very little, and variations in the phase amount to changing the seed number of the simulation.
\label{A:initialisation}
For a late-time initialisation of the scalar field, a relevant question is how late the initialisation can be made. At this point, we discovered that there is some pre-symmetry-breaking evolution that needs to be resolved. Before the symmetry breaking occurs globally, the density contrast in the simulation causes $|T|<\rho_*$ locally. The spatial Laplacian in the wave equation prevents the field from symmetry breaking in this region until the density has dropped below $\rho_*$ for a sufficiently large scale that seems to be related to the Compton wavelength $L_C$. However, during the time before this occurs, the domain structure establishes itself, even though the field amplitude is still very small (see figure \ref{fig:prevolution}). This evolution is relevant to resolve, and when we did not resolve it, the domain walls were more unstable in some instance{s}. Additionally, because the field is expected to be correlated on the Compton scale $L$, which increases before the symmetry breaking (see figure \ref{fig:timeresolution_rule}), we expect to have to initialise the simulation so that the field thermalises on this scale before the symmetry breaking. We chose both initialisation times $z_{\mrm{in}}=3,1$ in section \ref{S:hrsims} by running $512^3$ grid $(500 \,\mrm{Mpc/h})^3$ volume simulations, ensuring that we started the simulations before the domain structure established, and finding convergence with yet later initialisation times. We expect the time required before symmetry breaking to depend on the density contrasts, among other parameters, and we leave the analytic prediction of this for the future.

\FloatBarrier

\section{Solvers}\label{A:solvers}

\begin{SCfigure*}
    \centering
    \includegraphics[width=1.5\linewidth]{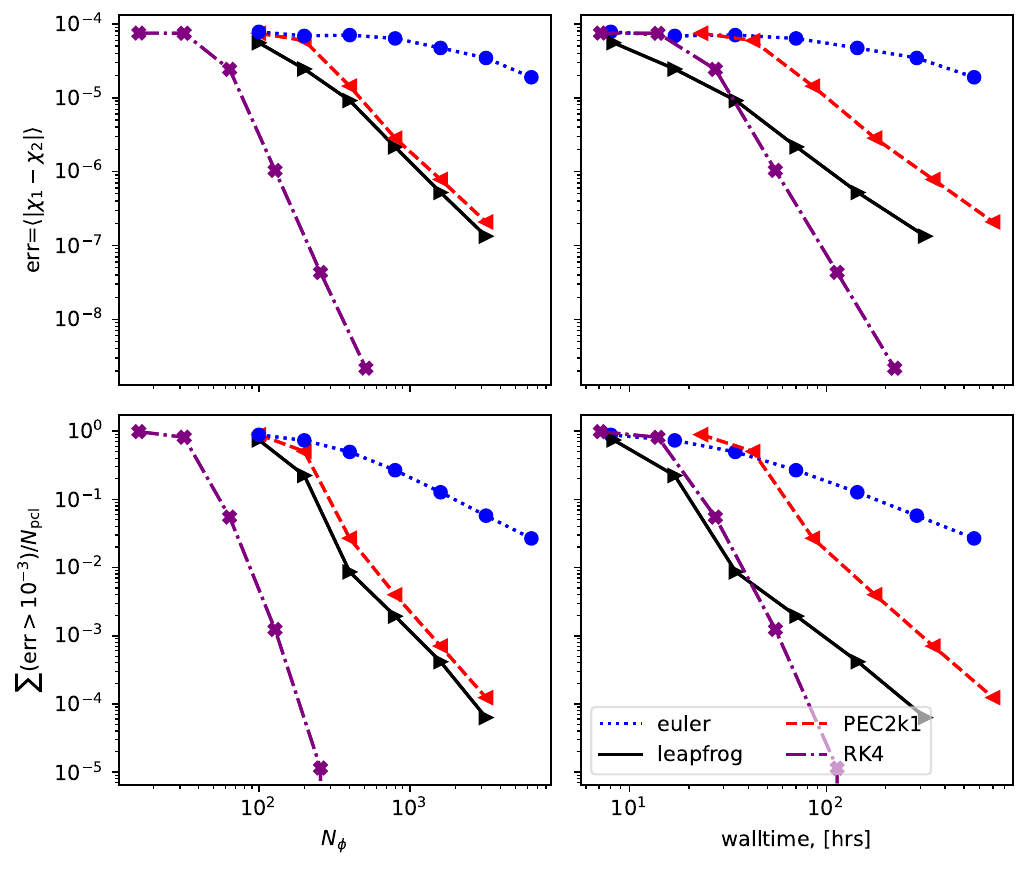}
    \caption{
        \protect\rule{0ex}{5ex}
    Comparison of different solvers for the symmetron in a $3$ Gpc/h box and $128^3$ grids. The convergence is checked in a redshift interval $z\in (1.5,1.46)$. The final slopes of the solver in logspace are 1.97 for the leapfrog,
    4.3 for the fourth-order Runge-Kutta, 1.91 for PEC2k1, and 0.87 for the Euler solver.
}
    \label{fig:solvers}
\end{SCfigure*}

 The solvers we chose were expensive in different ways. Using a method such as {the} fourth-order Runge-Kutta method required us to keep multiple auxiliary fields to avoid overwriting the field values before the neighbour lattice points had calculated their Laplacians. In addition, every partial step had to be done within its own loop over the lattice, giving more overhead. Finally, between each step, the auxiliary field that was used in the next loop Laplacian must have updated ghost layers of the computational domains before entering the loop, giving more communication overhead. Taking all of these obstacles into account, we compare  the convergence of both solvers with respect to the wall-clock time and with the number of iterations in figure \ref{fig:solvers}. The comparison is made over a redshift interval of $z\in (1.5,1.46)$ for the symmetron model parameters $(L_C,a_*,\beta_*) = (1\text{ Mpc/h}, 0.33,1)$. At each step $i$, the error was calculated as the average absolute difference between the solution obtained with a step size $h$ and the previous with step size $h/2$ as
 \begin{align}
     \mrm{err} \equiv \langle | \chi_i-\chi_{i-1}| \rangle.
  \end{align} 
Figure \ref{fig:solvers} shows the expected convergence with smaller step sizes. The Runge Kutta method is of the fourth order, meaning $\mrm{err}\sim\mathcal{O}(h^4)$. This easily recognised by realising that if $\mrm{err}\sim\mathcal{O}(h^n)$, then
 \begin{align}
    \chi_i(2h)-\chi(h) = \chi-\chi + \left[ \mrm{err}(2h)-\mrm{err}(h)\right] \sim (2^{n}-1)h^n,
 \end{align} 
so that the function in logspace is $\log(\mrm{err})\sim n\log h$. 

\begin{SCfigure*}
    \centering
    \includegraphics[width=1.2\linewidth]{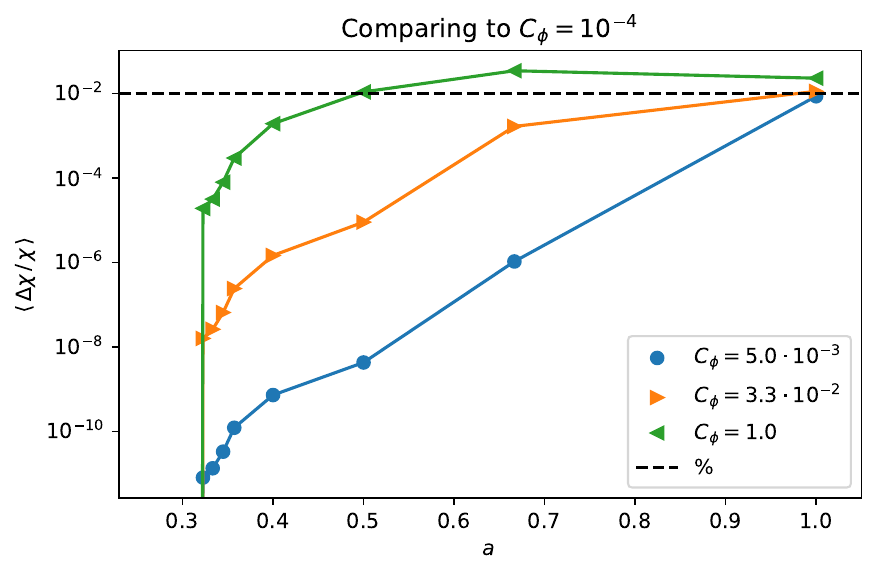}
    \caption{    \protect\rule{0ex}{5ex}
    Comparison of different Courant factors for the scalar field in a $512$ Mpc/h box and $128^3$ grids. The comparison is made with respect to the solution of the scalar field found with a Courant factor of $10^{-4}$. The percent-level agreement is indicated with the dashed horizontal line.}
    \label{fig:scalar_courant}
\end{SCfigure*}

In figure \ref{fig:scalar_courant} we demonstrate this convergence in the relative error of the actual scalar field value for a variety of choices of scalar Courant factors for cosmological time evolutions using the fourth-order Runge Kutta solver. Because of the divergence for $\chi\sim 0$, we only averaged the points on the lattice for which $\chi>0.1$.


\end{appendix}
\end{document}